\documentclass[
  journal=macro,
  manuscript=article,
  layout=traditional,
  spacingset=double
]{achemso}

\usepackage{chemformula} 
\usepackage[T1]{fontenc} 

\usepackage{hyperref}
\usepackage{times}
\usepackage{mathrsfs}
\usepackage{mathptmx}

\title{Machine Learning Framework for Characterizing Processing–Structure Relationship in Block Copolymer Thin Films}

\author{Bradley Lamb}
\affiliation{School of Polymer Science and Engineering, University of Southern Mississippi, 118 College Drive, Hattiesburg, MS 39406, USA}
\altaffiliation{These authors contributed equally to this work.}

\author{Saroj Upreti}
\affiliation{School of Polymer Science and Engineering, University of Southern Mississippi, 118 College Drive, Hattiesburg, MS 39406, USA}
\altaffiliation{These authors contributed equally to this work.}

\author{Yunfei Wang}
\affiliation{School of Polymer Science and Engineering, University of Southern Mississippi, 118 College Drive, Hattiesburg, MS 39406, USA}%

\author{Daniel Struble}
\affiliation{School of Polymer Science and Engineering, University of Southern Mississippi, 118 College Drive, Hattiesburg, MS 39406, USA}%

\author{Chenhui Zhu}
\affiliation{Advanced Light Source, Lawrence Berkeley National Laboratory, Berkeley, CA 94720, USA}%

\author{Guillaume Freychet}
\affiliation{NSLS-II, Brookhaven National
Laboratory, Upton, New York 11973, USA}
\altaffiliation{University of Grenoble Alpes, CEA, Leti, Grenoble, F-38000 France}

\author{Xiaodan Gu}
\affiliation{School of Polymer Science and Engineering, University of Southern Mississippi, 118 College Drive, Hattiesburg, MS 39406, USA}%
\email{xiaodan.gu@usm.edu}

\author{Boran Ma}
\affiliation{School of Polymer Science and Engineering, University of Southern Mississippi, 118 College Drive, Hattiesburg, MS 39406, USA}%
\email{boran.ma@usm.edu}


\begin{document}

\begin{abstract}
The morphology of block copolymers (BCPs) critically influences material properties and applications. This work introduces a machine learning (ML)-enabled, high-throughput framework for analyzing grazing incidence small-angle X-ray scattering (GISAXS) data and atomic force microscopy (AFM) images to characterize BCP thin film morphology. A convolutional neural network was trained to classify AFM images by surface features, achieving 97\% testing accuracy. Classified images were then analyzed to extract 2D grain size measurements from the samples in a high-throughput manner. ML models were trained to predict domain orientation based on processing parameters such as solvent ratio, additive type, and additive ratio. GISAXS-based properties were predicted with strong performances ($R^2$ > 0.75), while AFM-based property predictions were less accurate ($R^2$ < 0.60), likely due to the localized nature of AFM measurements compared to the bulk information captured by GISAXS. Beyond model performance, interpretability was addressed using Shapley Additive exPlanations (SHAP). SHAP analysis revealed that the additive ratio had the largest impact on morphological predictions, where additive provides the BCP chains with increased volume to rearrange into thermodynamically favorable morphologies. This interpretability helps validate model predictions and offers insight into parameter importance. Altogether, the presented framework combining high-throughput characterization and interpretable ML offers an approach to exploring and optimizing BCP thin film morphology across a broad processing landscape.
  
\end{abstract}

\section{Introduction}
Copolymer thin films consist of a wide range of materials with a vast array of applications spanning from membrane technologies for gas separation to energy applications including solid electrolytes.\cite{ghosal_gas_1994,mosciatti_light-modulation_2016,hou_applications_2019} They exhibit a diverse range of tunable properties dependent on both chemical structure and morphology. Copolymers, specifically block copolymers (BCPs), are a class of polymers in which multiple chemically distinct polymer chains, as building blocks, are bonded together forming one macromolecule. When the blocks of a copolymer are immiscible, unfavorable interactions between the blocks lead to phase separation resulting in unique morphologies.\cite{darling_directing_2007} These unique nanoscale phase-separated microdomains can be leveraged to facilitate the design of advanced materials.\cite{heeger_25th_2014,liu_unraveling_2018} For example, BCP membranes exhibit strong potential to surpass traditional organic and inorganic membranes by offering improved selectivity, permeability, and pore uniformity for applications such as water purification, battery electrolytes, and gas separation fields.\cite{wang_recent_2024}

BCP self-assembly behavior at thermodynamic equilibrium can be accurately predicted using self-consistent field theory.\cite{bates_block_1990,helfand_block_1976,leibler_theory_1980} The theory-predicted morphology of BCPs is primarily dictated by interactions between the blocks, known as the Flory-Huggins interaction parameter ${\chi}$, the degree of polymerization $N$, and the volume fraction of each block ${\phi}$. \cite{matsen_unifying_1996} In solid state, the phase diagram can increase in complexity due to polymer chains becoming kinetically trapped during sample preparation, resulting in a non-equilibrium state. Furthermore, the complexity increases in BCP thin films  due to interfacial energy and confinement effects as the morphology now depends on the strength of interaction at interfaces.\cite{mansky_controlling_1997,han_perpendicular_2009,brassat_nanoscale_2020}  

To achieve precise control over BCP solid state morphology, processing conditions such as solvent type, sample casting method, and post-deposition annealing technique require optimization.\cite{albalak_thermal_1997} The importance of solvent type is especially pronounced when the solvent preferentially swells one block, altering the volume fraction and thus the morphology.\cite{albalak_solvent_1998} Additionally, annealing techniques such as thermal and solvent-vapor annealing of BCP thin films play a large role in the final morphology. These techniques allow for increased mobility of the polymer chains, leading to a more thermodynamically favorable morphology.\cite{wang_recent_2024} In such cases, the kinetic effects of temperature change and solvent removal also play a key role in the final morphology.\cite{gu_situ_2014,albert_systematic_2012} Moreover, casting methods such as drop casting and spin coating of BCP solution also affect the morphology mainly due to variation in solvent evaporation dynamics. Spin coating promotes rapid solvent removal and shear induced alignment resulting in uniform films with well-ordered microdomains and reduced defect densities.\cite{zhang_influence_2012,tyona_theoritical_2013} On the other hand, slower solvent evaporation during drop casting leads to thicker films with varied domain orientation and rougher surface features.\cite{zhang_influence_2012,tyona_theoritical_2013,jung_effect_2010} Therefore, these differences affect overall phase behavior and domain spacing, as the interplay between solvent evaporation, interfacial energetics, and film thickness determine the degree of phase separation and ordering.\cite{jung_effect_2010,mishra_effect_2012,saito_mechanical_2021}  

Scattering and microscopy are common techniques for morphological characterization of BCP films. Grazing incidence small-angle X-ray scattering (GISAXS) is an inverse-space characterization technique measuring the phase behavior of BCP films to extract the average domain spacing and degree of ordering, where the domain of a BCP refers to the aggregation of a specific block.\cite{li_alternating_2019} Atomic force microscopy (AFM) is a real-space characterization technique that can provide insights in BCP surface morphology and serve as a complementary characterization method to GISAXS.\cite{lorenzoni_assessing_2015} Moreover, periodically ordered arrangements of BCP domains, i.e., grains, can be identified and measured in AFM as well. However, traditional analysis of BCP characterization data remains time consuming, as manual analysis is often required, especially for features such as grain size from AFM images. 

Despite yielding great achievements, the traditional process of BCP solid state morphology control is constrained by laborious experimentation and data analysis required from researchers for the optimization of processing conditions. Design of Experiments (DoE) has long been used to systematically reduce the experimental burden by probing a controlled subset of processing conditions and inferring trends.\cite{cao_how_2018} However, DoE typically assumes linear or low-order polynomial behavior, limiting its ability to model complex, nonlinear behavior and interactions, especially in high-dimensional or constrained experimental spaces.\cite{williamson_design_2022} Even with advanced DoE techniques such as high-dimensional DoE, these methods may still fail to capture the full complexity of response surfaces in systems with many variables and nonlinear responses.\cite{bukys_high-dimensional_2020} Furthermore, attempts to incorporate chemical identity into these frameworks by encoding reagents or materials as continuous variables, often through single-value descriptors, offer only a partial solution.\cite{wall_practical_2025} Because such representations compress inherently multidimensional chemical features into a single number, they struggle to preserve the true chemical variability, limiting the usefulness of DoE when comparing distinct chemistries.

In contrast, high-throughput data analysis and machine learning (ML) algorithms have demonstrated promising results in the field of materials science, offering more flexible approaches for capturing complex, multivariate relationships and solving optimization problems.\cite{struble_prospective_2024,upadhya_automation_2021,macleod_self-driving_2020} 
However, these algorithms typically require large datasets for the training process, which remains scarce for polymer systems, necessitating high-throughput experiments to generate the required data.\cite{chen_polymer_2021,patra_data-driven_2022,day_navigating_2023,schuett_application_2024} Therefore, high-throughput data acquisition and analysis, combined with ML-aided data interpretation, can be leveraged to explore the processing space for controlling the solid-state morphology of BCPs.  

ML algorithms are extremely diverse, tailored for specific tasks and datasets. These algorithms, in the case of supervised learning, extract relationships from a labeled datasets while optimizing a defined performance metric.\cite{jordan_machine_2015} Within supervised learning, there are two primary tasks: regression and classification. The key difference between these two tasks is the output: a numerical value for regression and a categorical output for classification.\cite{castillo-boton_machine_2022} In the field of materials science, these models have provided researchers with an approach to reducing experimentation by predicting the properties of untested systems, as well as a tool to understand the relationships within complex material systems.\cite{wei_machine_2019,ma_machine-learning-assisted_2023,lu_accelerated_2018,oliynyk_high-throughput_2016}
Numerous models have been applied in the materials science field to great success, including random forest (RF), support vector machine (SVM), multilayer perceptron (MLP), and extreme gradient boosting (XGB).\cite{fridgeirsdottir_multiple_2018,tao_benchmarking_2021,kim_comparison_2021,banerjee_mathematical_2022,liang_machine-learning-assisted_2021,xu_machine_2021,chen_predicting_2021,xu_accurate_2008} 

Within the polymer field, ML has been leveraged to decipher scattering profiles and AFM images of polymers, elucidating polymer conformational and morphological features.\cite{tung_small_2022, ding_deciphering_2025, kobayashi_machine_2023, paruchuri_machine_2024} Specifically, the Jayaraman group has pioneered the integration of ML and molecular simulations to advance the interpretation of scattering data from complex polymer solutions. Their approach enables the analysis of scattering profiles that lack conventional fitting equations, overcoming a major limitation in characterizing novel polymer architectures. \cite{beltran-villegas_computational_2019, wu_machine_2022, wessels_machine_2021}Furthermore, Doerk and coworkers demonstrated an autonomous platform for the discovery of novel BCP morphologies through a directed self-assembly approach. Their work leveraged a gaussian processor to guide the direction of future scattering experiments to further study these novel BCP morphologies.\cite{doerk_autonomous_2023} These works significantly broaden the applicability of scattering experiments, offering new insights into the structure and behavior of emerging polymer systems.

Despite the emerging use of ML for polymer property predictions in the solution and solid state, comparatively little work has focused on the utilization of ML to understand how processing conditions influence the solid-state morphology of BCPs -- an area critical for the design of functional materials. Furthermore, generating large datasets in the field of polymer science remains a significant challenge due to long experimentation and analysis times. Many studies have circumvented this issue by compiling pre-existing data from online databases and published papers into a single dataset.\cite{ethier_predicting_2022,kim_polymer_2021,sha_machine_2021,struble_prospective_2024} However, this compiled data can be highly variable, as it comes from different laboratories, leading to inconsistencies in both the experimental setups and data analysis, as well as missed or mislabeled descriptors. 

\begin{figure*}
\includegraphics[width=14 cm]{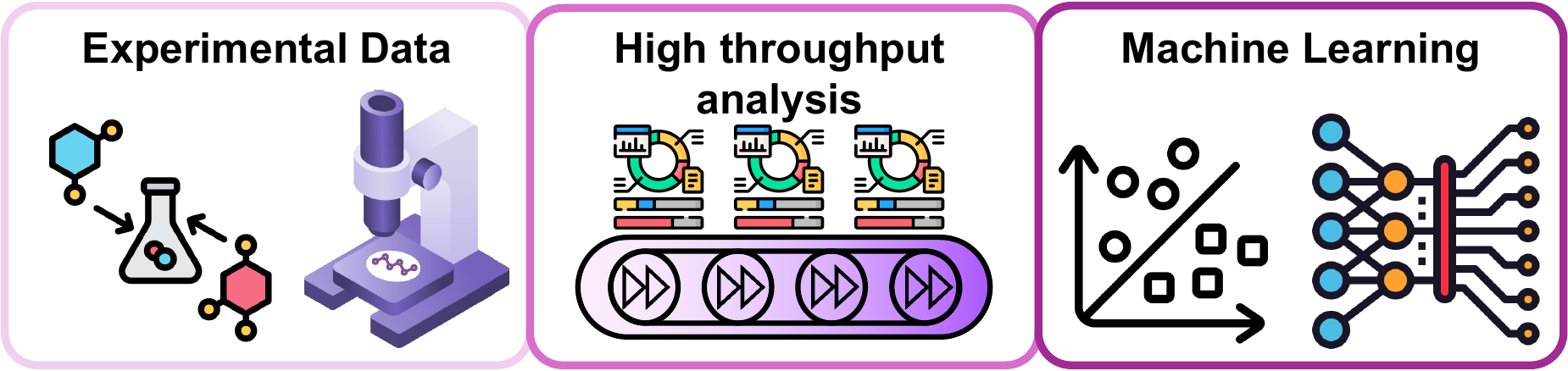}
\caption{\label{fig:intro} Process flow for the high-throughput analysis of experimentally collected data, curated for machine learning applications and property predictions.}
\end{figure*}

Therefore, the synergistic use of high-throughput data analysis and ML methods can address the previously mentioned challenges and accelerate the rational design of polymeric materials.\cite{eyke_toward_2021, beaucage_automation_2024} In this work, we focus on a single block copolymer system, polystyrene-block-poly(ethylene oxide) (PS-\textit{b}-PEO, Mn = 23-\textit{b}-7 kg/mol), which forms hexagonally packed cylindrical microdomains. This well-characterized composition serves as a model to investigate how processing conditions can drive transitions in domain orientation, as well as changes in ordering and domain size. We present an approach for the high-throughput data analysis of BCP thin film GISAXS profiles and two-dimensional grain sizes measured by AFM, creating a robust dataset of morphological properties from over 200 BCP thin films. In addition, interpretable ML is leveraged to evaluate the relationship between processing parameters and BCP morphology, and offers direct morphological property predictions from processing conditions. This work serves as a proof of concept for an integrated high-throughput data analysis-ML framework for the optimization of processing conditions for polymeric materials (Fig. \ref{fig:intro}).

\section{Methodology}

\subsection{Experimentation and characterization}

Polystyrene-block-poly(ethylene oxide) (PS-\textit{b}-PEO) (Mn = 23-\textit{b}-7 kg/mol), which forms a hexagonally packed cylinder morphology, was used as a model system to generate thin films using the “nonvolatile additive solvent annealing” (NVASA) approach. The details of the NVASA process can be found in our previous publication.\cite{weller_roll--roll_2019} PS-\textit{b}-PEO was dissolved overnight in host solvents composed of mixtures of toluene and tetrahydrofuran (THF) at the following volume ratios, 100:0, 90:10, 80:20, 70:30, 60:40 and 50:50, to form 2 to 3 wt\% stock solutions. With NVASA, it has been previously reported that a controlled amount of high boiling point additive can briefly remain in the polymer films after the primary low boiling point solvent evaporates after spin coating, thereby swelling the BCP thin film, providing segmental mobility and improving domain order. \cite{shui_rapid_2025} Therefore, for each stock solution, varied masses of a high boiling point additive: chloronaphthalene (CN), or methylnaphthalene (MN), were added to achieve additive ratios from 1.00 to 7.00. Despite the differing amounts of additive, the rate of solvent removal has been shown to remain relatively consistent.\cite{weller_roll--roll_2019} The control of additive ratio is dictated by the mass of additive relative to the mass of BCP (Equation \ref{eq:add_ratio}).
\begin{eqnarray}
\label{eq:add_ratio}
\text{Additive Ratio}=\frac{mass_{additive}}{mass_{BCP}}+1
\end{eqnarray}

For instance, an additive ratio of 1.00 indicates no additive (therefore no swelling of the BCP film), whereas additive ratios of 2.00, 3.00, and 3.25 equate to 1:1, 2:1, and 2.25:1 mass ratio of additive:BCP, respectively. These solutions were then spin coated at 2000 rpm for 60 s on silicon wafers after oxygen plasma cleaning for 5 min (Diener Inc. at 10 mTorr, 20 sccm O2, 40 W). The coated films were dried under ambient conditions. 202 samples were prepared in total.  

GISAXS experiments were carried out at beamline 7.3.3 of Advanced Light Source (ALS) at Lawrence Berkeley National Laboratory, and Beamline 12-ID of National Synchrotron Light Source II (NSLS-II) at Brookhaven National Laboratory. The X-ray energies used at the ALS and NSLS-II were 10 keV and 16.1 keV, respectively. The incidence angle between the sample surface and X-rays was 0.14° and 0.12° for ALS and NSLS-II respectively, and the sample-to-detector distance was 3.5 meters and 4 meters at the ALS and NSLS-II, respectively. The scattering profiles were recorded on Pilatus 2M detector (ALS) and a Pilatus 1M detector (NSLS-II). Igor Pro software, equipped with the Nika and Irena packages, was used to process the 2D scattering patterns and extract 1D line profiles, which were subsequently analyzed using a high-throughput approach.

Oxford Cypher Asylum AFM was used to image the domain orientation of BCP thin films. Phase and height images (1 $\mu$m $\times$ 1 $\mu$m) with resolution of 256 $\times$ 256 pixels were taken in tapping mode. The tapping mode cantilevers (Tap300AI-G) with force constant of 40 N/m and resonant frequency of 300 kHz were used to scan the surface of films at a rate of 1.95 Hz. AFM raw data were collected using Igor Pro–based software, then flattened, and exported as images using Gwyddion software. These images were later processed in high-throughput fashion to extract properties of interest. 

\subsection{High-throughput data analysis}
\subsubsection{GISAXS}

\begin{figure}
\includegraphics[width=6.5 cm]{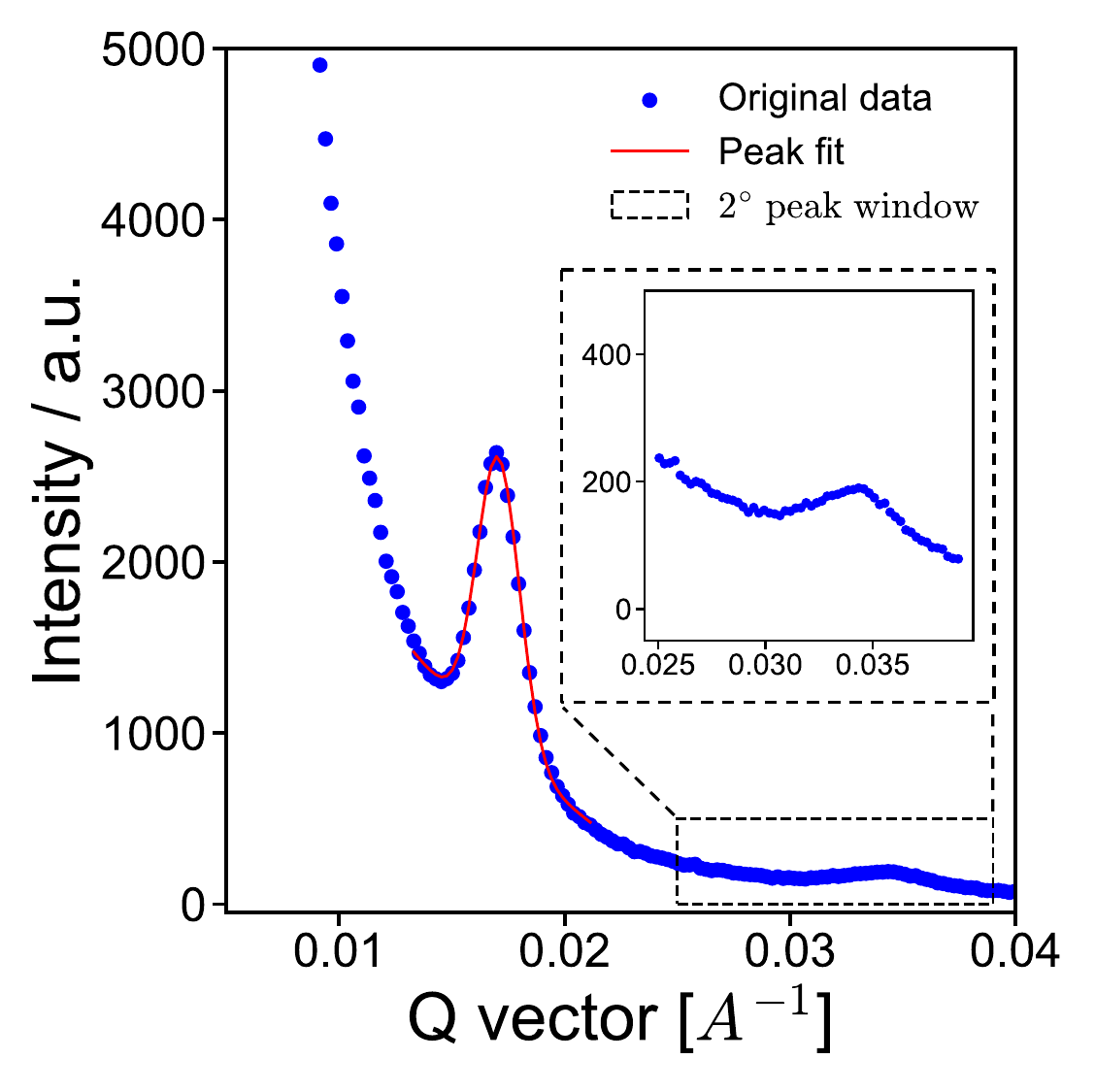}
\caption{\label{fig:fitting} Example peak fitting of 1D GISAXS profile, restricting the fitting functions within the peak region, indicated by the dashed box, to improve fitting of the scattering peak related to the BCP sample.}
\end{figure}

The 1D GISAXS profiles, obtained from horizontal line cuts of the 2D scattering patterns, were analyzed by first identifying and isolating the primary peak, which is related to the characteristic length scale in the sample, or domain spacing. The peak region was then fit to a combination of a power-law decay and a Gaussian function (Equation \ref{eq:one}):
\begin{eqnarray}
\label{eq:one}
I(q) = A + Cq^{-d} + D e^{\frac{-4\ln(2)(q - b)^2}{w^2}}
\end{eqnarray}
In this equation, $I(q)$ represents the scattering intensity as a function of the scattering vector $q$. The constant $A$ accounts for a baseline background offset, while the term $Cq^{-d}$ models a power law decay commonly associated with background or form factor scattering, with $C$ as the amplitude and $d$ as the decay exponent. The Gaussian peak is characterized by an amplitude $D$, a center position $b$, and a full width at half maximum (FWHM) $w$. The peak position $b$ is used to calculate the domain spacing $ds$, using the relation $ds=2\pi/b$. The FWHM, $w$, reflects the distribution of domain spacings and serves as an indicator of the degree of structural order in the sample. This fitting function enables simultaneous characterization of both periodic structural features and background scattering behavior.\cite{gu_situ_2014} An in-house Python script was used to fit the scattering data by first locating the primary peak position, then using this value as a starting parameter for the fitting equation, which the Levenberg-Marquardt optimization algorithm was used to minimize the sum of squared residuals. The optimization process was allowed a maximum of 50,000 function evaluations, i.e., the model could be evaluated up to 50,000 times with different parameter values during the search for the optimal fit. The script outputs the final fitted values for peak position and FWHM, along with the coefficient of determination ($R^2$) of the fitted function. This process for GISAXS analysis is both highly accurate and computationally efficient, with an average fitting $R^2$ of 0.99, taking less than 1 second per sample (Fig. \ref{fig:fitting}). 

To verify that the domain orientations observed in AFM were consistent with the bulk thin film structure, higher-order peaks were analyzed relative to the primary peak. In the scattering profiles, higher-order peaks appeared only in samples with horizontally-oriented cylindrical domains. However, the expected (11) peak at a spacing ratio of \(\sqrt{2}\) was absent; instead, the (20) peak was observed at a ratio of 2. This behavior is typical for horizontally-oriented hexagonal cylinders, where the structure factor suppresses the (11) peak.\cite{bendejacq_well-ordered_2002, chu_morphologies_1995} Because a ratio of 2 is also consistent with vertically-stacked lamellae, the spacing ratio alone could not confirm the bulk morphology. To resolve this ambiguity, 2D GISAXS patterns were analyzed (Fig. S1), confirming the presence of horizontally-oriented hexagonal cylinders rather than vertical lamellae.\cite{smarsly_quantitative_2005} These patterns are discussed in detail in the SI.

Automated peak ratio detection was carried out by isolating the region between 1.5 and 2.5 times the q-position of the primary peak, in this case corresponding to the expected (20) peak for hexagonally packed cylinders (Fig. \ref{fig:fitting}). This window can be adjusted to target peaks from other morphologies. The algorithm then scans the region for a secondary peak. If none is found, no output is saved. If a peak is detected, the ratio is calculated and stored, provided it falls within a user-defined range. Otherwise, the output file flags the peak as outside the expected range, indicating further analysis is needed. Using this approach, secondary peaks were correctly identified with 92\% accuracy.

\subsubsection{AFM}

\begin{figure*}
\includegraphics[width=16cm]{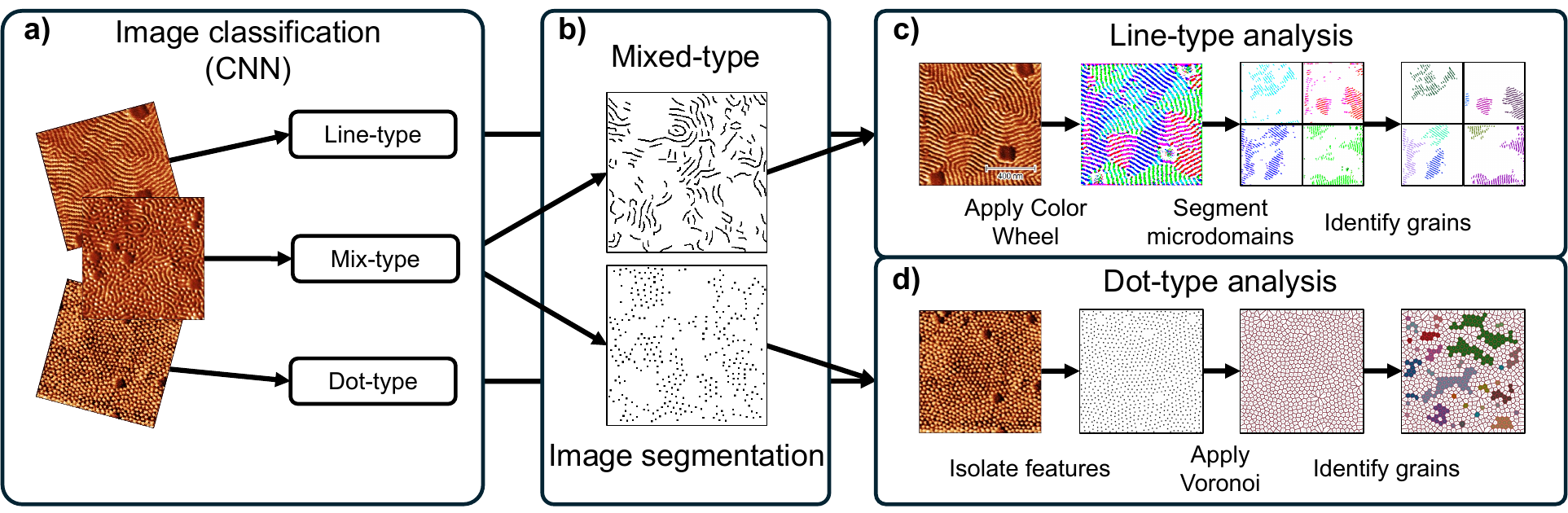}
\caption{\label{fig:afm_proc} Overview of the morphology agnostic, feature-based AFM image analysis process for extracting grain size. a) AFM images are classified based on domain orientation by a convolutional neural network (CNN). b) Images containing mixed-type features are segmented into dot- and line-type features. c) Color-wheel analysis of line-type images, color-coding microdomains based on orientation, then segmenting the color-coded microdomains before determining grains. d) Voronoi analysis of dot-type images, features are first isolated before applying the Voronoi analysis and identifying grains.}
\end{figure*}

The BCP thin films in this study are expected to arrange in hexagonal cylinder morphologies, given the PS-b-PEO block ratio. However, the orientation and ordering of the cylinders is highly dependent on the processing conditions of the BCP films. \cite{hao_self-assembly_2017} Therefore, the 202 AFM images collected represent various orientations of hexagonally packed cylinders, including vertical, horizontal, and a coexistence of both orientations, which manifest as `dot'-, `line'-, and `mixed'-type features in the AFM images, respectively. While no single approach is sufficient to analyze the local ordering, or grain size, of the microdomains for different orientations, multiple approaches can be used in tandem to create a robust AFM analysis approach (Fig. \ref{fig:afm_proc}). Furthermore, this approach is not limited to characterizing hexagonally packed cylinder morphology, as is performed in this study, and could be extended to other morphologies that similarly yield 'dot'- and 'line'-type surface features.

AFM images are first classified into three categories (Fig. \ref{fig:afm_proc} a), in this study depending on the orientation of the hexagonal cylinders, with the use of a convolutional neural network (CNN), discussed in more detail in the following section. After classification, the binarized dot- and line-type images were ready for analysis, however, the mixed-type images required further pre-processing. Segmentation of the mixed-type images was performed using the labelme software to create two distinct masks for the dot- and line-type features (Fig. \ref{fig:afm_proc} b). \cite{wada_wkentarolabelme_2021} The segmented masks were treated the same as the the dot- and line-type images for further analysis. 

For line-type images, the MLExchange color wheel was applied to the Two-dimensional fast Fourier transforms (2D FFT) of the AFM images, which assigns a unique color to the microdomains of different orientations (Fig. \ref{fig:afm_proc} c).\cite{zhao_mlexchange_2022} After isolating one phase from the color wheel image, i.e. filtering out one of the blocks, the microdomains were segmented by creating masks related to each orientation. Once isolated, the scikit-learn cluster package was used to group adjacent pixels together, which were then filtered according to a size threshold to reduce noise in the images.\cite{pedregosa_scikit-learn_2018} Candidate grains were identified by another clustering analysis before applying another threshold to filter out grains smaller than 150\% of the average, again to reduce noise and artificial grains. The choice of this threshold was determined empirically, with a range of values tested, and 150\% consistently producing reasonable results across all samples (Fig. S2), discussed in more detail in the SI. The final grain size was then calculated by averaging the grain sizes for each sample, then converting from pixels$^2$ to $\mu$m$^2$. In total, this grain analysis process takes approximately 8 seconds per image. Furthermore, the in-house grain analysis code was evaluated, and results were generally in agreement with grain analysis from a separate study,\cite{murphy_automated_2015} (Fig. S3) which is discussed in more detail in the SI (Table S1). 

Separately, the dot-type images were analyzed using Voronoi analysis (Fig. \ref{fig:afm_proc} d). Traditionally, Voronoi type analyses are used to identify grain boundaries and defects in well-ordered BCP systems.\cite{hammond_adjustment_2003, bourne_hexagonal_2014} The following approach, however, leverages the Voronoi analysis to characterize two-dimensional grain sizes for weakly-ordered systems. Within the analysis, the dot-type features were first isolated using the labelme software, creating discrete points for each dot-type feature. Voronoi cells were constructed by treating each feature in the image as a center point. The image was divided such that each cell contains the area closer to its feature than to any other, resulting in a set of perfectly adjacent, non-overlapping regions—one surrounding each dot-like feature. For an ideal hexagonal cylinder morphology, each Voronoi cell should have 6 equal edges, reflecting 6 surrounding hexagonally packed cylinders.\cite{hamley_hexagonal_1993} Therefore, the numbers of edges for each cell were determined, additionally having to satisfy a criteria being within a 55\% threshold of the binned mode edge length in the image, in order to filter edges at the end of the image or near a defect. Furthermore, for a cell to be considered for grain analysis, the lengths of its 6 edges must differ by no more than 2.5\% of the mean edge length of the cell, promoting the selection of well-ordered hexagonally packed structures. Finally, adjacent cells were grouped together, and the average grain size determined across all grains for a given sample and converted from pixels to pixels$^2$ to $\mu$m$^2$. The Voronoi analysis occurs rapidly, taking only 3 seconds per image.

Additionally, local domain spacing measurements were obtained from the AFM images. 2D FFTs were performed on the cleaned and binarized images to obtain the characteristic length scale of the phase-separated microstructure in the BCPs. The radial distribution profile was extracted based on the frequency spectrum in the Fourier space. The radial distribution data was then fit to Equation \ref{eq:one}, similar to the GISAXS data fitting, to improve resolution and more accurately determine the peak position than would be possible from the unfitted data alone. The correlation peak from the 2D FFT was then converted using Equation \ref{eq:two} to obtain the characteristic length scale (domain spacing, $ds$) in real space:
\begin{eqnarray}
\label{eq:two}
ds =\frac{FOV}{u}
\end{eqnarray}
where $u$ is the peak position of the radial distribution, and $FOV$ is the field of view of the AFM images (i.e. image size, such as 1 $\mu$m$^2$ for images in Fig. \ref{fig:afm_proc}).\cite{carmona_structure_2021} This process occurs rapidly, taking less than 10 seconds for all 202 samples, while maintaining a high degree of fitting accuracy with an $R^2$ of 0.96 with respect to the raw GISAXS data. 

\subsubsection{Convolutional Neural Network AFM Image Classification}

A CNN was first employed to classify the AFM images based on their surface features, specifically, line-, dot-, and mixed-type features, corresponding to vertically aligned cylinders, horizontally aligned hexagonal cylinders, and a combination of both. The CNN was trained using the binarized images as inputs and manually labeled classes as targets. Additionally, due to a significant imbalance of images with line-type features relative to the images with dot- and mixed-type features, the images within the latter classes were augmented, creating three copies of each image rotated in increments of 90°, creating a more balanced dataset. Furthermore, the dot-type images were reflected horizontally and vertically to further increase the amount of data for the class. After augmentation, the dataset was much more balanced, with 171, 84, and 72 images for line-, dot-, and mixed-type features, respectively. 

\begin{figure*}
\includegraphics[width=16cm]{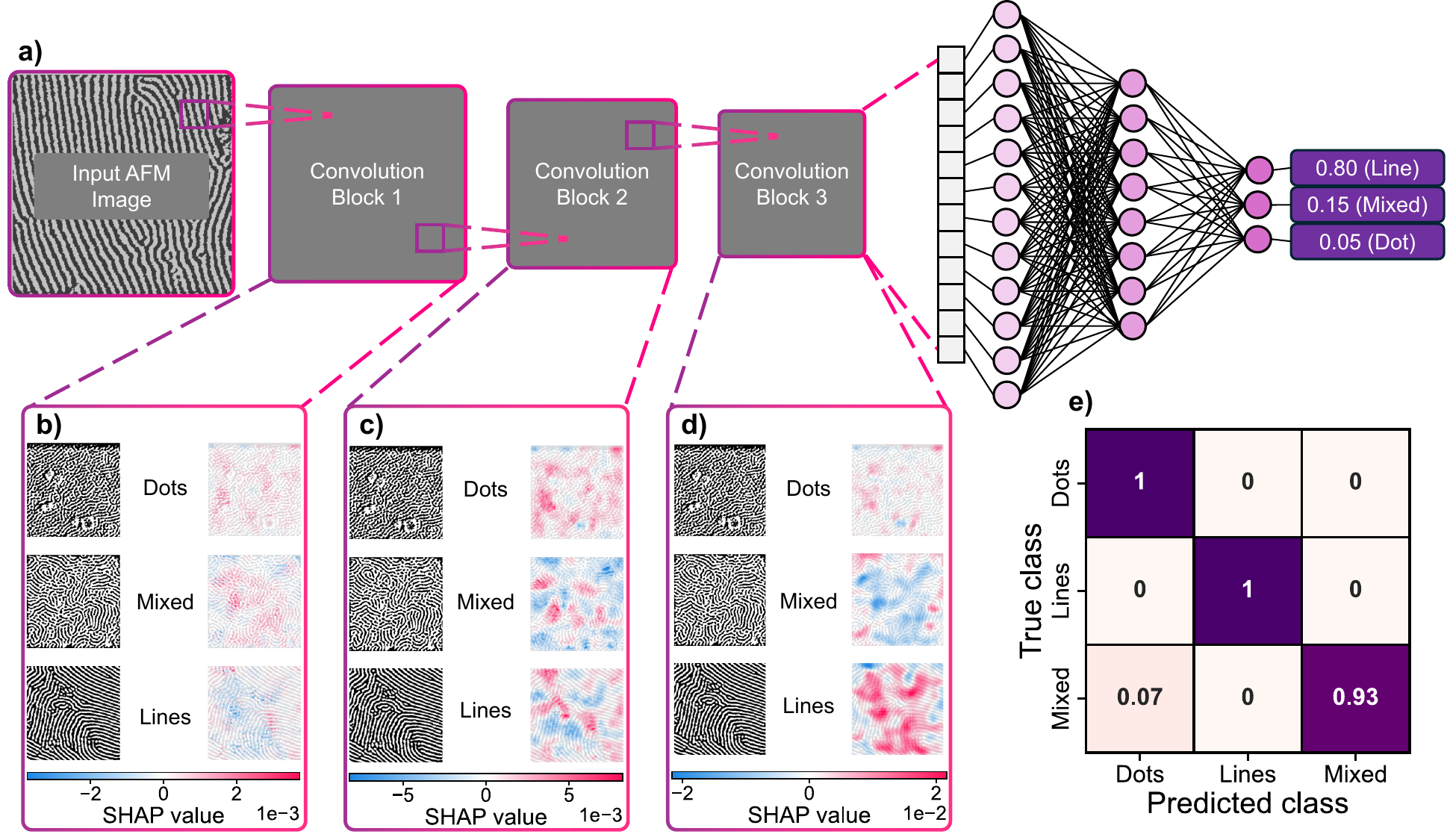}
\caption{\label{fig:cnn} a) Simplified architecture of the CNN used for surface feature classification. AFM images are passed through a series of convolutional blocks to extract hierarchical feature maps, which are then fed into fully connected layers to produce a probabilistic prediction over the three feature-based classes (line, dot, and mixed). b-d) SHAP-based interpretation of the CNN’s predictions at convolutional blocks 1-3. Positive SHAP values (red) highlight features that positively contribute to the predicted class, while negative SHAP values (blue) indicate features that negatively influence the prediction. e) Confusion matrix summarizing the CNN's performance on the test set.}
\end{figure*}

The model follows a simplified VGG-like structure,\cite{simonyan_very_2015} consisting of three convolutional blocks with two convolutional layers each, followed by batch normalization, ReLU activations, and max pooling (Fig. \ref{fig:cnn} a). A more detailed description of the CNN can be found in the SI. Among the 327 AFM images used in the classification task, 228 were allocated for training, 65 for validation, and 34 for testing. On the test set, the CNN model achieved a high classification accuracy of 97.06\%. To interpret the model’s decision-making process, the Shapley Additive exPlanations (SHAP) Python package was employed (Fig. \ref{fig:cnn} b-d). SHAP analysis provides a unified framework for explaining ML model predictions by assigning each feature an importance value, its SHAP value, based on its contribution to the model’s output.\cite{lundberg_unified_2017} When SHAP analysis is applied to a CNN trained on images, each pixel (or small region of pixels) is assigned a value that reflects how much it contributed to the prediction, relative to a reference baseline. Regions with positive contributions are shown in red, indicating they increase confidence in the predicted class, while regions with negative contributions are shown in blue, indicating they decrease confidence. In practice, this means that the SHAP maps highlight the specific parts of an image the CNN relies on. For example, in dot-type images, the dot clusters appear as red regions, showing that they are the main features driving the classification.

Within each convolutional block, features were extracted from the input to generate feature maps. After each block, spatial dimensions were reduced, causing fine-scale features to be captured in the earlier blocks, while progressively larger and more abstract features were analyzed in subsequent blocks. The SHAP analysis of block 1 reveals that features evaluated at smaller length scales exhibit strong positive correlations with dot- and mixed-type classifications, such as dots and disconnections in the line features. This aligns with the physical interpretation of the images, where these features characteristically lack prominent large structural features. Similarly, the SHAP analysis of block 2 reveals positive correlation between clusters of dots with the dot-type classification, and groups and lines with heavy curvature with the mixed-type classification. Interestingly, block 2 also begins to pick up parallel groups of line features, positively correlating them with the line-type classification. Finally, features extracted from block 3 primarily correlate positively with the line-type classification, focusing on long stretches of line features. Overall, the CNN performed exceptionally well distinguishing between dot- and line-type classifications, however, struggled with the mixed-type classifications, over-predicting the dot-type classification for some mixed-type images (Fig. \ref{fig:cnn} e).

\section{Results and Discussion}
\label{sec:rnd}
\subsection{Dataset overview}
\begin{figure*}
\includegraphics[width=16 cm]{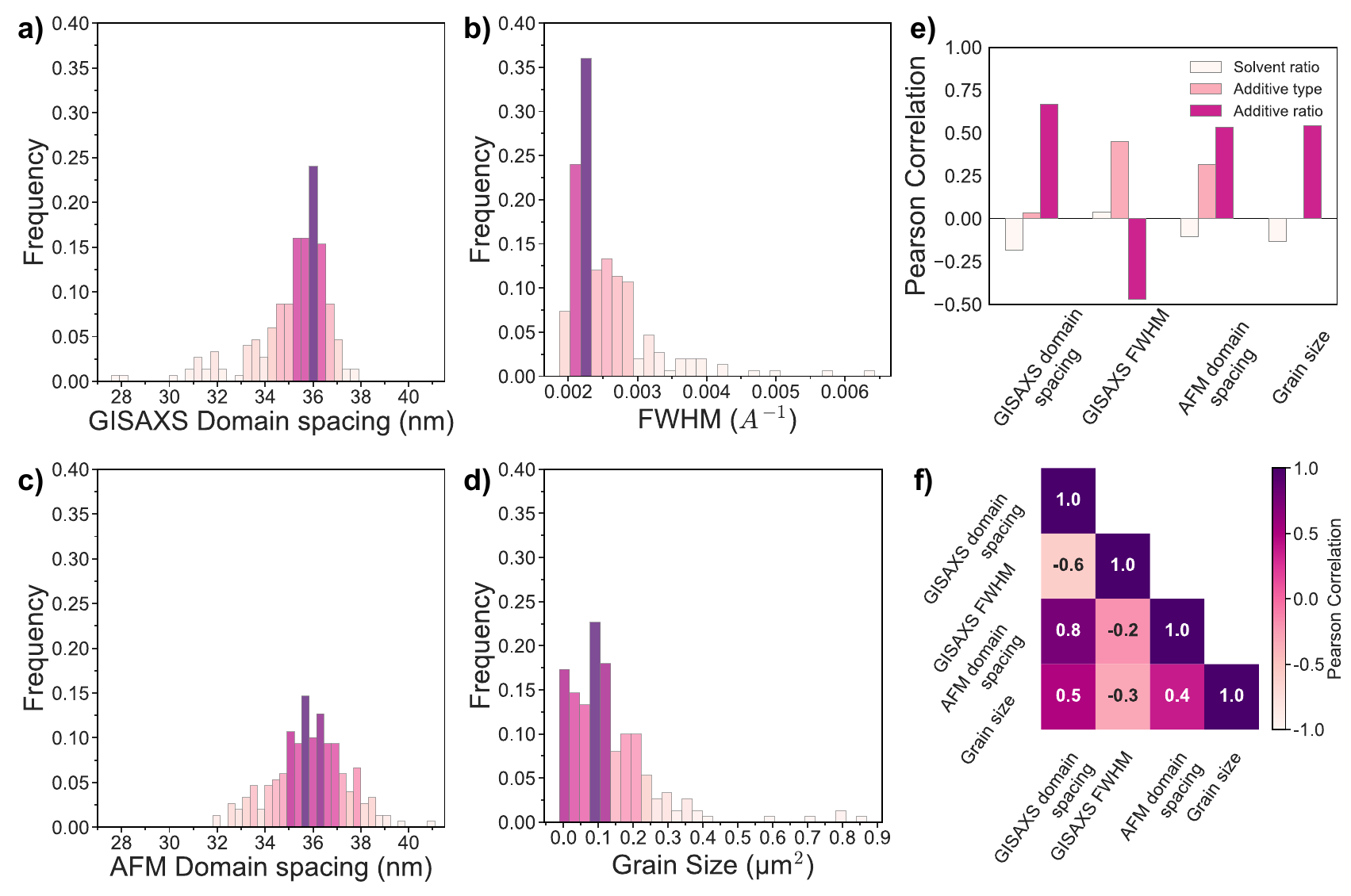}
\caption{\label{fig:dist} Frequency plots of a) GISAXS-measured domain spacing, b) GISAXS-measured FWHM, c) AFM-measured domain spacing, and d) AFM-measured grain size from the 202 BCP thin films. e) Pearson correlations for each processing condition input with respect to each measured output, and f) Pearson correlation heatmap of measured outputs.}
\end{figure*}

As mentioned previously, the processing parameters explored in this study include solvent ratios of toluene:THF from 0.5 to 1.0, additive ratios from 1.00 to 7.00, and additive types of CN and MN. Domain spacing and FWHM from GISAXS measurements show narrow distributions, with domain spacing measurements around 36 nm and FWHM around 0.0025 A$^{-1}$ (Fig. \ref{fig:dist} a, b). These distributions indicate the majority of the BCP films have relatively good ordering, indicated by the low FWHM value which is reflective of a narrow distribution of domain spacings in the sample. However, the AFM-measured morphological properties, domain spacing and grain size, demonstrate broader distributions (Fig. \ref{fig:dist} c, d). While the range of values for the GISAXS- and AFM-based measurements are in agreement with each other, the localized nature of the AFM measurements leads to higher variability, in contrast to GISAXS measurements which reflect the bulk properties of the sample.

Pearson correlation coefficients were calculated to assess potential linear relationships between the processing parameters and the resulting morphological properties (Fig. \ref{fig:dist} e). Among all processing parameters, the additive ratio exhibited the strongest Pearson correlation with the measured morphological properties. A previous study using \textit{in situ} ellipsometry showed that it takes a few minutes for the high boiling point additive to evaporate from the BCP films.\cite{ogieglo_situ_2015} During this period, the additive-infused BCP film exhibits enhanced segmental mobility, allowing polymer chains to rearrange into more ordered and thermodynamically favorable morphologies. Interestingly, the type of additive exhibits a relatively strong Pearson correlation with both the GISAXS-measured FWHM and AFM-measured domain spacing. For AFM, this may be attributed to surface-specific effects introduced by the additive, such as differential evaporation dynamics, not captured by bulk GISAXS measurements. In the case of GISAXS-measured FWHM, the correlation may reflect broader distributions of domain spacings within the sample, suggesting increased structural heterogeneity without significantly altering the average domain spacing. Meanwhile, solvent ratio has a relatively low Pearson correlation across all measured outputs. Although solvent plays a critical role in the initial solubilization of the polymers, it has little effect on the resulting morphology of the BCP film, due to the low boiling point solvent volatilizing during the spin coating process. Furthermore, the Pearson correlations were evaluated for the morphological outputs with respect to each other (Fig. \ref{fig:dist} f). The domain spacing from both GISAXS and AFM has a fairly large Pearson correlation, indicating good agreement between the measurements from the two approaches. Domain spacing from GISAXS and AFM are inversely correlated with GISAXS-measured FWHM, indicating that as the samples become more ordered, i.e., decreasing FWHM, domain spacing increases. Additionally, the grain size has positive Pearson correlations with domain spacing and an inverse correlation with FWHM, again indicating that as the samples become more ordered, the grain size increases. Scatter plots of the morphological properties are shown in the SI (Fig. S4).  

\subsection{Domain orientation classification} 
Classification models were trained using solvent ratio, additive ratio, and additive type as input features, with `dot-', `line-', and `mixed-type' classifications, based on AFM images, as the target output. This approach allows for the creation of a domain orientation map as a function of the processing conditions used to create the sample. Support Vector Classifier (SVC) and Random Forest Classifier (RFC) were selected for this task due to their robust performance and high accuracy in classification problems. The data set was split using 80\% as the training set, and 20\% for the test set to evaluate the model performance while using a 5-fold cross validation to ensure generalizability. A grid search approach was employed to evaluate various combinations of hyperparameters, which can be found in the SI for each model. The PyTorch accuracy metric was used to evaluate the models, in which the model with the best cross-validation accuracy was selected. 

\begin{figure*}
\includegraphics[width=16 cm]{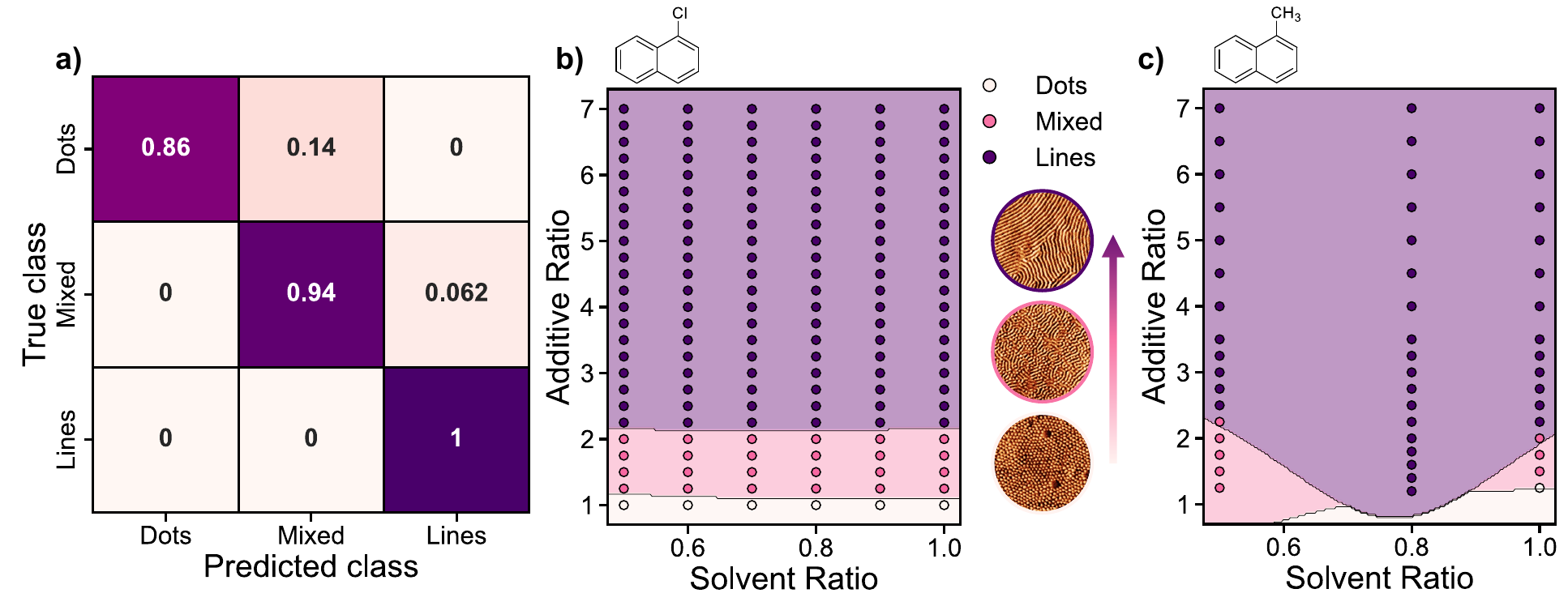}
\caption{\label{fig:o/d} a) Confusion matrix showing the testing results of the SVC model. b) Decision boundary plots illustrating the SVC decision boundaries as the transparent regions and the actual domain orientation types as dots. Separate plots were created for the CN additive, left, and the MN additive, right, as a function of solvent and additive ratios. For both additive types, increasing additive ratio promotes a transition from dot-type surface features to line-type surface features, \textit{i.e.} a transition from vertically oriented domains to horizontally oriented domains as the sample approaches thermodynamic equilibrium.}
\end{figure*}

Both classifiers performed exceedingly well, with RFC performing slightly better than SVC, with cross-validation accuracies of 96\% and 93\%, respectively. Although the RFC model outperformed in overall accuracy, its tree-based architecture yields rigid, stepwise phase boundaries due to poor interpolation between data points. In contrast, SVC defines smoother, continuous boundaries by maximizing inter-class margins, making it more suitable for constructing phase maps. Consequently, SVC was chosen for further analysis.
The SVC model overpredicted the line- and mixed-type surface features (Fig. \ref{fig:o/d} a), likely because the dataset was skewed towards line- (163) and mixed-type (32) features with few samples exhibiting dot-type (7). Two distinct phase maps were generated: one for samples with CN as the additive and another for MN. The CN map showed no misclassifications (Fig. \ref{fig:o/d} b, left), as the phase transitions were largely dependent on the additive ratio. In contrast, the MN map was more complex, with three misclassifications near phase boundaries (Fig. \ref{fig:o/d} b, right), likely a consequence of MN’s lower boiling point and greater impact on the solvent ratio. Despite these discrepancies, both systems exhibited a clear trend in which increasing additive ratio promotes a transition from dot-type (vertically stacked cylinders) to line-type (horizontally aligned cylinders) as the kinetic window is extended, allowing the samples to approach their thermodynamic equilibrium state.

\subsection{Processing-structure relationship prediction}

\begin{figure*}
\includegraphics[width=16cm]{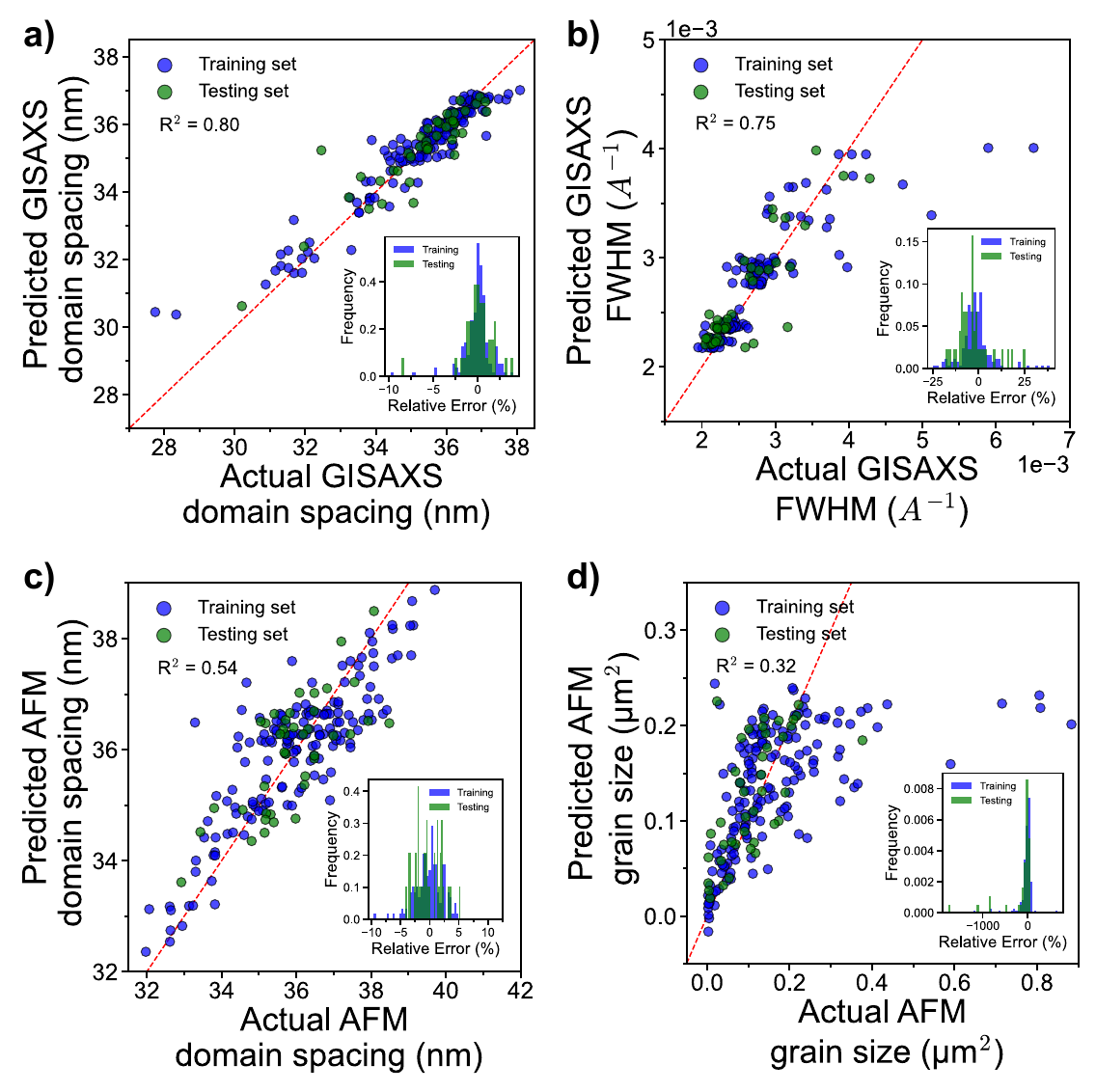}
\caption{\label{fig:models} Parity plots with relative error insets for regression model predictions: a) GISAXS-measured domain spacing and b) FWHM predicted by the RF models, and c) AFM-measured domain spacing and d) grain size predicted by the XGB and SVM models respectively.}
\end{figure*}

Four traditional ML models, including RF, SVM, MLP, and XGB, were trained on the dataset to predict specific morphological properties of the BCP films as a function of the processing conditions. The performances of the ML models were compared with more straightforward approaches, including linear, lasso, and Bayesian ridge regression. Results indicated that the ML models consistently outperformed the other approaches across all target properties (Fig. S5), discussed in more detail in the SI. Training was performed with a 5-fold cross-validation and grid-search, using 80\% of the data for training and 20\% for testing. The set of hyperparameters evaluated within the grid-search can be found in the SI. The evaluation metric used was the coefficient of determination ($R^2$) in which the model with the best average cross-validation $R^2$ was chosen to ensure generalizability (Fig. \ref{fig:models}). Furthermore, the effect of training set size was evaluated to ensure models could not be trained to achieve comparable performance on smaller datasets (Fig. S6).

RF performed the best for predicting GISAXS-measured domain spacing, achieving a testing $R^2$ of 0.80, with the majority of prediction errors falling within \textpm 2.5\% (Fig. \ref{fig:models} a). The relatively high accuracy is expected given the low variability and narrow distribution of the domain spacing data measured by GISAXS. However, the model performs much better in the high domain spacing region, around 36 nm, where the majority of the data is located, than in the low domain spacing region where data is much more sparse. Similarly, RF outperformed the other models when predicting the FWHM with a testing $R^2$ of 0.75 with the majority of error falling within \textpm 25\% (Fig. \ref{fig:models} b). The model performs exceptionally well when predicting FWHM within the `narrow-FWHM' regime, around 0.0025 A$^{-1}$, but struggled with predictions in the `broad-FWHM' region, where data was not as available. Despite the strong correlation and general agreement between the GISAXS and AFM domain spacing measurements, the models trained on domain spacing measured by AFM did not perform as well compared to the models trained on domain spacing measured by GISAXS. XGB performed the best with a testing $R^2$ of 0.54, with the majority of error distributed within \textpm 5\% (Fig. \ref{fig:models} c). Despite the better distribution of data relative to the GISAXS-measured domain spacing, the high variability of the data likely affected the model's ability to accurately predict the measurements. Interestingly, SVM was the best performing model for predicting the grain size with a testing $R^2$ of 0.32, but with a much more significant error spread than the previous models (Fig. \ref{fig:models} d). The relatively poor accuracy is again likely due to the local nature of the AFM measurements being highly variable across a given sample. Overall, models trained to predict GISAXS-measured properties outperformed those trained on AFM-measured properties, likely because GISAXS captures bulk characteristics, while AFM probes more localized surface features, leading to increased variability within the data. Additional results from each model can be found in the SI (Figs. S7-S10).

 The RF models trained on GISAXS-measured domain spacing and FWHM were further evaluated with SHAP given their relatively high degree of accuracy to interpret the model's predictions and evaluate feature importance.\cite{lundberg_local_2020} The mean SHAP value for each processing parameter reflects its average influence on the model's predictions, e.g. domain spacing or FWHM. Local SHAP values, by contrast, describe the influence of a parameter on an individual sample's prediction. A positive SHAP value indicates the parameter drives the predicted property higher relative to the model's average prediction, while a negative SHAP value drives the predicted property to lower values.

\begin{figure}
\includegraphics[width=16cm]{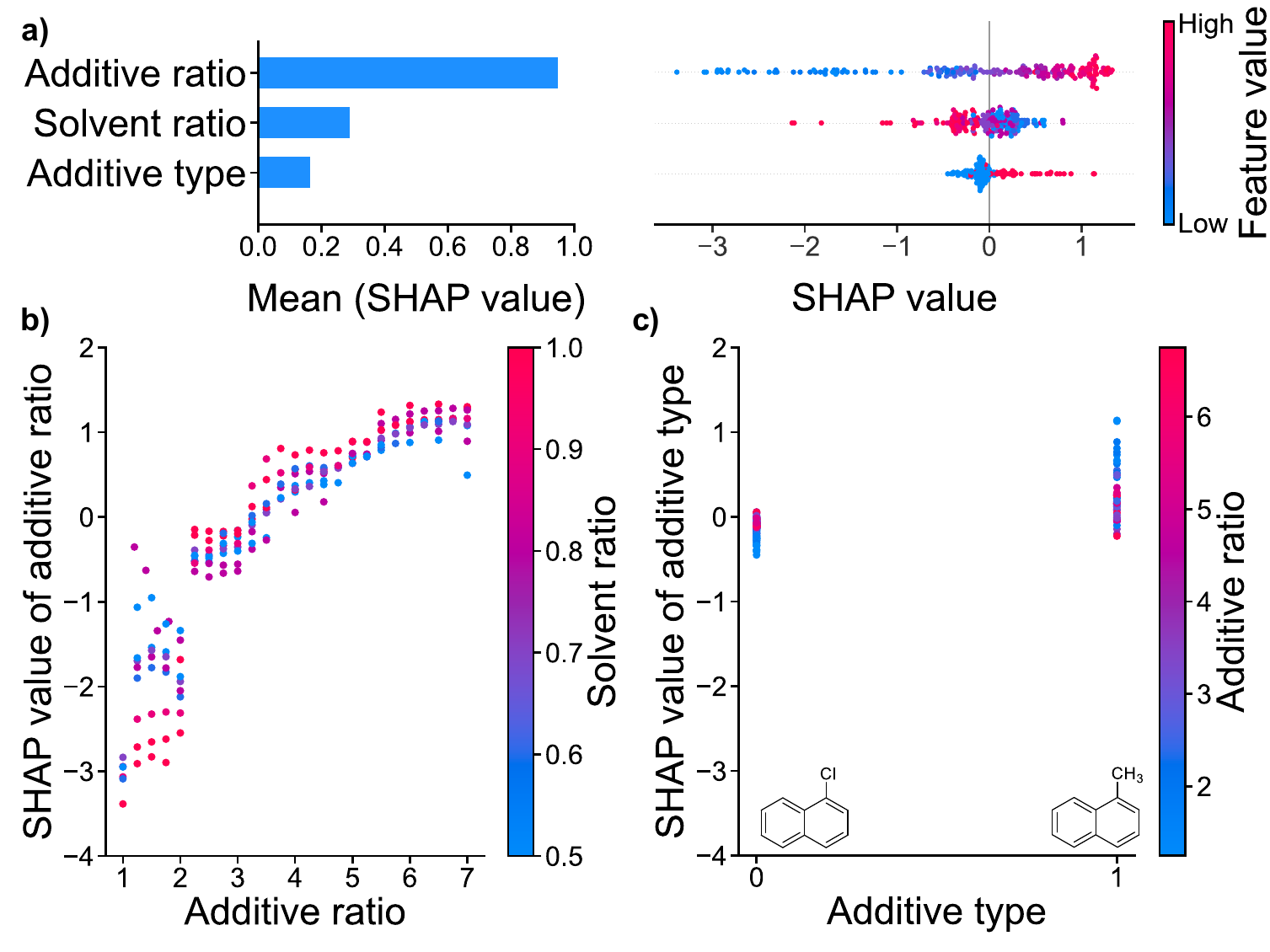}
\caption{\label{fig:shap_ds} SHAP plots from RFR model trained on GISAXS-measured domain spacing. a) Left: bar chart of the mean SHAP value. Right: beeswarm plots of SHAP value corresponding to individual data points where the dots position on the x-axis indicates the influence of the feature on the model's prediction for the given data point. SHAP dependence plots of b) additive ratio and solvent ratio and c) additive type and additive ratio.}
\end{figure}

For the RF model trained on GISAXS-measured domain spacing data, additive ratio has a much larger mean SHAP value, 0.95, than solvent ratio, 0.29, and additive type, 0.17 (Fig. \ref{fig:shap_ds} a left). This indicates that across all samples, additive ratio has an influence greater than 3 times that of the other processing parameters on the resulting BCP domain spacing. This is because the low-volatility additive extends the temporal window for the BCP to rearrange into thermodynamically favorable morphologies, which otherwise remain kinetically trapped. Due to the relatively high volatility of the solvents, their effects on domain spacing are marginal relative to additive ratio, which is reflected by the substantially lower mean SHAP values. Interestingly, despite small changes in volatility between the additives, the additive type has only a minor overall influence on the predictions, as indicated by its low mean SHAP value.

Moreover, analysis of the local SHAP values for each feature provides a more detailed view of how the processing parameters influence domain spacing predictions on a per-sample basis. Low values of additive ratio (represented by blue dots) have a large spread of SHAP values while higher values of additive ratio (red dots) aggregate around a central SHAP value (Fig. \ref{fig:shap_ds} a right). This suggests that domain spacing predictions are more sensitive to changes at lower additive ratios, with the influence tapering off as the additive ratio increases. Although the SHAP values for solvent ratio generally cluster around zero, higher solvent ratios, corresponding to increased toluene content, tend to have a negative impact on the domain spacing predictions, leading to smaller predicted values. This indicates that, although the solvents volatilize during spin coating, the processing history from the solvents have some effects on the final domain spacing of the BCP films. The SHAP values for additive type 0 (CN) are tightly clustered below zero, indicating that this additive type has a minimal and consistent influence on domain spacing, with little variability across samples. However, for additive type 1 (MN) the SHAP values are much more dispersed, reflecting greater variability in domain spacing when using MN relative to CN. 

Furthermore, the local SHAP values were plotted with respect to processing conditions to identify potential cross-relationships between the parameters (Fig. \ref{fig:shap_ds} b, c and Fig. S11). At low additive ratios, low values of solvent ratio generally lead to slightly higher predicted domain spacing, corresponding to well-ordered samples (Fig. \ref{fig:shap_ds} b). This suggests co-solvent systems, solubilizing each segment of the BCP result in better ordered BCP morphologies at low additive ratios. Additionally, as shown in Fig. \ref{fig:shap_ds} c, the SHAP values for additive type reveal that the influence of CN on predictions of domain spacing are narrowly aggregated, indicating CN provides good control over BCP morphology. On the other hand, the SHAP values for MN are much more dispersed, and slightly larger than those for CN. Given the slight increase in volatility of MN relative to CN, the increased mobility of the chains does not last as long for MN, therefore kinetically trapping the BCP chains in potentially unfavorable arrangements, leading to more variable, unpredictable domain spacings.

\begin{figure}
\includegraphics[width=16cm]{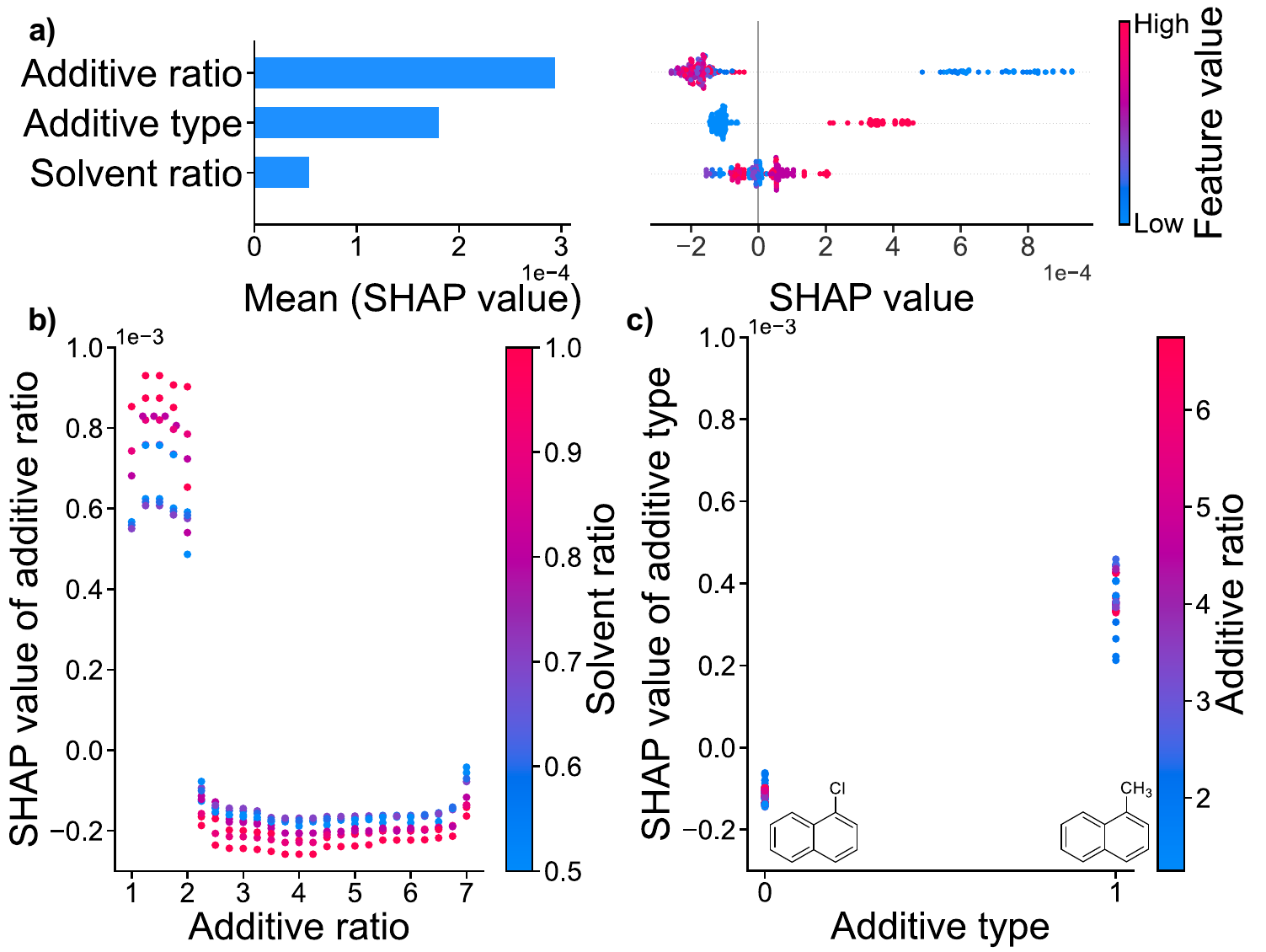}
\caption{\label{fig:shap} SHAP plots from RFR model trained on GISAXS-measured FWHM. a) Left: bar chart of the mean SHAP value. Right: beeswarm plots of SHAP value corresponding to individual data points where the dots position on the x-axis indicates the influence of the feature on the model's prediction for the given data point. SHAP dependence plots of b) additive ratio and solvent ratio and c) additive type and additive ratio.}
\end{figure}

While the SHAP evaluations for the GISAXS-measured FWHM and domain spacing models share some similarities, key differences also emerge. Interestingly, the additive type has a much more significant mean SHAP value for the FWHM predictions, indicating the additive type has a more profound impact on BCP FWHM than domain spacing (Fig. \ref{fig:shap} a left). However, additive ratio still maintains the largest mean SHAP value, 0.00029, followed by additive type, 0.00018, and solvent ratio, 0.000054. Similar to the local SHAP analysis for domain spacing, high additive ratios cluster at lower SHAP values, reflecting rapid ordering of the BCP, whereas lower additive ratios exhibit larger and more scattered SHAP values, indicative of weakly-ordered samples with broad FWHM. (Fig. \ref{fig:shap} a right). Although the SHAP values for additive type show a similar trend for those related to the domain spacing model, they are far more distinctly separated in the FWHM model, indicating substantial variation in FWHM with changes of additive type. Diving deeper into the SHAP analysis for the FWHM prediction, an interesting relationship can be observed between solvent ratio and additive ratio (Fig. \ref{fig:shap} b and Fig. S12). Similarly to the trend observed in the domain spacing model (Fig. \ref{fig:shap_ds} b) but more pronounced, low solvent ratios, reflecting a mixture of toluene and THF, are associated with lower FWHM predictions at low additive ratios, indicating improved ordering from the co-solvents improving the mobility of both blocks in the BCP solution. However, at higher additive ratios, i.e., greater than 2, the relationship reverses and lower solvent ratios contribute to higher FWHM predictions. This suggests a complex relationship between the solvent processing history, the additive ratio, and the FWHM of the BCP films. Additionally, when plotting the SHAP values of additive type against the additive type itself, CN (type 0) shows tightly clustered contributions to FWHM, suggesting consistent and controlled effects on BCP film ordering (Fig. \ref{fig:shap} c). In contrast, MN (type 1) is associated with broader and higher FWHM predictions, indicating more variability and poorer control. These results suggest that small changes in volatility of the additive could lead to measurable changes in the ordering of the BCP films.

\section{Conclusion}

In summary, a framework enabling the high-throughput analysis of BCP thin film morphology was developed, facilitating analysis of GISAXS and AFM characterization data. The process for extracting domain spacing from both GISAXS and AFM measurements is highly accurate, time efficient, and provides higher resolution values than the raw data provides. Average fitting $R^2$ values of 0.99 and 0.96 were achieved for the GISAXS and AFM data, respectively, taking less than one second per sample. Additionally, higher-order peaks were observed in the GISAXS data, and the determination of the spacing ratio between the peaks was automated with an accuracy of 92\%. Furthermore, two-dimensional values of grain size were obtained by utilizing the AFM image analysis approach of classification, segmentation, and color-wheel or Voronoi analysis depending on the surface feature type. The CNN classifier proved to be highly accurate and computationally inexpensive, while providing physically accurate predictions reflected by the SHAP analysis. Overall, this provides a platform for the rapid exploration of BCP morphological space by accelerating the rate of morphological analysis for the BCP films. 

Additionally, the rapid morphological analysis of the BCP films enabled the creation of a dataset for ML model training. SVM models were highly capable of accurately classifying sample domain orientation when given processing conditions, while a CNN achieved similar accuracy using AFM images directly as inputs. RF models were extremely accurate when predicting morphological properties measured from GISAXS, however, precision was limited when predicting below 34 nm due to an imbalance of data. Meanwhile, XGB and SVM models trained on AFM data were generally accurate but their predictive capabilities suffered from variability within the AFM data. SHAP analysis was employed to interpret the processing–-structure relationships learned by the RF models from the GISAXS data, confirming that the model’s predictions were physically meaningful. Beyond validating the strong influence of additive ratio, SHAP also revealed interactions between processing history, specifically the solvent ratio and additive ratio, and the GISAXS--measured domain spacing and FWHM. Despite the similar boiling points of the two additives, the slightly lower boiling point of MN resulted in decreased domain spacing, increased FWHM, and more variability among samples relative to those with CN, indicating slight changes in the volatility of the additive have pronounced effects on BCP thin film morphology. The discussed framework combining high-throughput morphological analysis with interpretable ML has demonstrated highly accurate evaluation of BCP thin film morphology. Furthermore, the results not only provided a predictive tool for BCP thin film morphology based on processing conditions, but also a quantitative understanding of the processing--structure relationships. 

Future studies could benefit from this approach and focus on expanding the design space to include new polymer chemistries, molecular weights, and different solvent types. This would enable analysis of a broader range of block copolymer morphologies, including gyroid and other complex structures, beyond the hexagonal cylinder morphology studied here. Such an expansion would strengthen the framework allowing for analysis of all BCP morphologies. Additionally, the variation in AFM data could be greatly improved by collecting multiple replicates across each sample, as well as increasing the field of view of the images. This would not only decrease the variability of the data, but also increase the size of grains able to be measured, allowing for investigation of much more well-ordered samples. Moving forward, these changes should result in significant improvements in the performance of ML models trained on AFM data. However, collection of AFM images is currently one of the largest bottlenecks in this process, requiring significant experimentation time. Increasing both the number of AFM images and the size of the images would further exasperate this bottleneck. Therefore, utilizing a super-resolution approach could allow for the rapid collection of low-resolution AFM images which could be transformed into high resolution space.\cite{chang_accelerating_2019} With the above-mentioned recommendations in mind, the exploration of BCP thin film morphologies could be greatly accelerated with the use of such an ML-enabled, integrated high-throughput data analysis approach. 

\begin{acknowledgement}

The authors acknowledge the support from DOE Grant DE-SC0024432. B.M. thanks the Director of the School of Polymer Science and Engineering, the Dean of the College of Arts and Sciences, and the Vice President for Research, all at the University of Southern Mississippi, for their support with generous start-up funds. This research used beamline 7.3.3 of the Advanced Light Source, which is a DOE Office of Science User Facility under contract no. DE- AC02-05CH11231. Y.W. was supported in part by an ALS Doctoral Fellowship in Residence. This research used the Soft Matter Interfaces (SMI) beamline at the National Synchrotron Light Source II, a U.S. DOE Office of Science User Facility operated by Brookhaven National Laboratory under contract no. DE-SC0012704. X.G. thanks NSF for partially supporting his work under the grant no. OIA-2229686.

\end{acknowledgement}

\section{Data Availability}

The dataset and the code used in this study are available at \href{https://github.com/MaResearchLab/BCP-ML-Characterization-Framework}{https://github.com/MaResearchLab/BCP-ML-Characterization-Framework}.

\section{Supporting Information}

The Supporting Information is available free of charge at {Our Paper Link}. This file includes details of GISAXS bulk morphology analysis; validation and verification of AFM analysis methods; CNN architecture; morphological datasets derived from GISAXS and AFM; machine learning model comparisons and grid-search parameter space; the effect of training-set size on model performance; morphology predictions from all trained models; and SHAP interaction plots.

\bibliography{References.bib}

@article{ghosal_gas_1994,
	title = {Gas separation using polymer membranes: an overview},
	volume = {5},
	copyright = {Copyright © 1994 John Wiley \& Sons, Ltd.},
	issn = {1099-1581},
	shorttitle = {Gas separation using polymer membranes},
	url = {https://onlinelibrary.wiley.com/doi/abs/10.1002/pat.1994.220051102},
	doi = {10.1002/pat.1994.220051102},
	language = {en},
	number = {11},
	urldate = {2024-07-02},
	journal = {Polymers for Advanced Technologies},
	author = {Ghosal, Kanchan and Freeman, Benny D.},
	year = {1994},
	note = {\_eprint: https://onlinelibrary.wiley.com/doi/pdf/10.1002/pat.1994.220051102},
	pages = {673--697},
}

@article{darling_directing_2007,
	title = {Directing the self-assembly of block copolymers},
	volume = {32},
	issn = {0079-6700},
	url = {https://www.sciencedirect.com/science/article/pii/S0079670007000627},
	doi = {10.1016/j.progpolymsci.2007.05.004},
	number = {10},
	urldate = {2024-07-02},
	journal = {Progress in Polymer Science},
	author = {Darling, S. B.},
	month = oct,
	year = {2007},
	pages = {1152--1204},
}

@article{heeger_25th_2014,
	title = {25th {Anniversary} {Article}: {Bulk} {Heterojunction} {Solar} {Cells}: {Understanding} the {Mechanism} of {Operation}},
	volume = {26},
	copyright = {© 2013 WILEY-VCH Verlag GmbH \& Co. KGaA, Weinheim},
	issn = {1521-4095},
	shorttitle = {25th {Anniversary} {Article}},
	url = {https://onlinelibrary.wiley.com/doi/abs/10.1002/adma.201304373},
	doi = {10.1002/adma.201304373},
	language = {en},
	number = {1},
	urldate = {2024-07-02},
	journal = {Advanced Materials},
	author = {Heeger, Alan J.},
	year = {2014},
	note = {\_eprint: https://onlinelibrary.wiley.com/doi/pdf/10.1002/adma.201304373},
	pages = {10--28},
}

@article{mosciatti_light-modulation_2016,
	title = {Light-{Modulation} of the {Charge} {Injection} in a {Polymer} {Thin}-{Film} {Transistor} by {Functionalizing} the {Electrodes} with {Bistable} {Photochromic} {Self}-{Assembled} {Monolayers}},
	volume = {28},
	copyright = {© 2016 WILEY-VCH Verlag GmbH \& Co. KGaA, Weinheim},
	issn = {1521-4095},
	url = {https://onlinelibrary.wiley.com/doi/abs/10.1002/adma.201600651},
	doi = {10.1002/adma.201600651},
	number = {31},
	urldate = {2024-07-02},
	journal = {Advanced Materials},
	author = {Mosciatti, Thomas and del Rosso, Maria G. and Herder, Martin and Frisch, Johannes and Koch, Norbert and Hecht, Stefan and Orgiu, Emanuele and Samorì, Paolo},
	year = {2016},
	note = {\_eprint: https://onlinelibrary.wiley.com/doi/pdf/10.1002/adma.201600651},
	pages = {6606--6611},
}

@article{wang_recent_2024,
	title = {Recent progress on fabrication and applications of advanced block copolymer membranes},
	volume = {39},
	issn = {2214-9937},
	url = {https://www.sciencedirect.com/science/article/pii/S2214993724000356},
	doi = {10.1016/j.susmat.2024.e00855},
	urldate = {2024-07-02},
	journal = {Sustainable Materials and Technologies},
	author = {Wang, Xue-Qi and Wang, Tao and Feng, Ying-Nan and Zhang, Lu-Yao and Zhao, Zhi-Ping},
	month = apr,
	year = {2024},
	pages = {e00855},
}

@article{bates_block_1990,
	title = {Block {Copolymer} {Thermodynamics}: {Theory} and {Experiment}},
	volume = {41},
	issn = {0066-426X, 1545-1593},
	shorttitle = {Block {Copolymer} {Thermodynamics}},
	url = {https://www.annualreviews.org/content/journals/10.1146/annurev.pc.41.100190.002521},
	doi = {10.1146/annurev.pc.41.100190.002521},
	language = {en},
	number = {Volume 41,},
	urldate = {2024-07-02},
	journal = {Annual Review of Physical Chemistry},
	author = {Bates, Frank S. and Fredrickson, Glenn H.},
	month = oct,
	year = {1990},
	note = {Publisher: Annual Reviews},
	pages = {525--557},
}

@article{helfand_block_1976,
	title = {Block {Copolymer} {Theory}. 4. {Narrow} {Interphase} {Approximation}},
	volume = {9},
	issn = {0024-9297},
	url = {https://doi.org/10.1021/ma60054a001},
	doi = {10.1021/ma60054a001},
	number = {6},
	urldate = {2024-07-02},
	journal = {Macromolecules},
	author = {Helfand, Eugene and Wasserman, Z. R.},
	month = nov,
	year = {1976},
	note = {Publisher: American Chemical Society},
	pages = {879--888},
}

@article{han_perpendicular_2009,
	title = {Perpendicular {Orientation} of {Domains} in {Cylinder}-{Forming} {Block} {Copolymer} {Thick} {Films} by {Controlled} {Interfacial} {Interactions}},
	volume = {42},
	issn = {0024-9297},
	url = {https://doi.org/10.1021/ma9002903},
	doi = {10.1021/ma9002903},
	number = {13},
	urldate = {2024-07-02},
	journal = {Macromolecules},
	author = {Han, Eungnak and Stuen, Karl O. and Leolukman, Melvina and Liu, Chi-Chun and Nealey, Paul F. and Gopalan, Padma},
	month = jul,
	year = {2009},
	note = {Publisher: American Chemical Society},
	pages = {4896--4901},
}

@article{albalak_thermal_1997,
	title = {Thermal annealing of roll-cast triblock copolymer films},
	volume = {38},
	issn = {0032-3861},
	url = {https://www.sciencedirect.com/science/article/pii/S003238619600938X},
	doi = {10.1016/S0032-3861(96)00938-X},
	number = {15},
	urldate = {2024-07-02},
	journal = {Polymer},
	author = {Albalak, Ramon J. and Thomas, Edwin L. and Capel, Malcolm S.},
	month = jul,
	year = {1997},
	pages = {3819--3825},
}

@article{albalak_solvent_1998,
	title = {Solvent swelling of roll-cast triblock copolymer films},
	volume = {39},
	issn = {0032-3861},
	url = {https://www.sciencedirect.com/science/article/pii/S0032386197004977},
	doi = {10.1016/S0032-3861(97)00497-7},
	number = {8},
	urldate = {2024-07-02},
	journal = {Polymer},
	author = {Albalak, Ramon J. and Capel, Malcolm S. and Thomas, Edwin L.},
	month = jan,
	year = {1998},
	pages = {1647--1656},
}

@article{zhang_influence_2012,
	title = {Influence of film casting method on block copolymer ordering in thin films},
	volume = {8},
	issn = {1744-6848},
	url = {https://pubs.rsc.org/en/content/articlelanding/2012/sm/c2sm07308k},
	doi = {10.1039/C2SM07308K},
	language = {en},
	number = {18},
	urldate = {2024-07-02},
	journal = {Soft Matter},
	author = {Zhang, Xiaohua and Douglas, Jack F. and Jones, Ronald L.},
	month = apr,
	year = {2012},
	note = {Publisher: The Royal Society of Chemistry},
	pages = {4980--4987},
}

@article{tyona_theoritical_2013,
	title = {A theoritical study on spin coating technique},
	volume = {2},
	url = {https://doi.org/10.12989/AMR.2013.2.4.195},
	doi = {10.12989/AMR.2013.2.4.195},
	language = {en},
	number = {4},
	urldate = {2024-07-02},
	journal = {Advances in materials Research},
	author = {Tyona, M.D.},
	month = dec,
	year = {2013},
	pages = {195--208},
}

@article{jung_effect_2010,
	title = {Effect of film thickness on the phase behaviors of diblock copolymer thin film},
	volume = {4},
	issn = {1936-0851},
	url = {https://tohoku.elsevierpure.com/en/publications/effect-of-film-thickness-on-the-phase-behaviors-of-diblock-copoly},
	doi = {10.1021/nn1003309},
	language = {English},
	number = {6},
	urldate = {2024-07-02},
	journal = {ACS Nano},
	author = {Jung, Jueun and Park, Hae Woong and Lee, Sekyung and Lee, Hyojoon and Chang, Taihyun and Matsunaga, Kazuyuki and Jinnai, Hiroshi},
	month = jun,
	year = {2010},
	pmid = {20499924},
	note = {Publisher: American Chemical Society},
	pages = {3109--3116},
}

@article{mishra_effect_2012,
	title = {Effect of {Film} {Thickness} and {Domain} {Spacing} on {Defect} {Densities} in {Directed} {Self}-{Assembly} of {Cylindrical} {Morphology} {Block} {Copolymers}},
	volume = {6},
	issn = {1936-0851},
	url = {https://doi.org/10.1021/nn205120j},
	doi = {10.1021/nn205120j},
	number = {3},
	urldate = {2024-07-02},
	journal = {ACS Nano},
	author = {Mishra, Vindhya and Fredrickson, Glenn H. and Kramer, Edward J.},
	month = mar,
	year = {2012},
	note = {Publisher: American Chemical Society},
	pages = {2629--2641},
}

@article{saito_mechanical_2021,
	title = {Mechanical {Properties} of {Ultrathin} {Polystyrene}-b-{Polybutadiene}-b-{Polystyrene} {Block} {Copolymer} {Films}: {Film} {Thickness}-{Dependent} {Young}’s {Modulus}},
	volume = {54},
	issn = {0024-9297},
	shorttitle = {Mechanical {Properties} of {Ultrathin} {Polystyrene}-b-{Polybutadiene}-b-{Polystyrene} {Block} {Copolymer} {Films}},
	url = {https://doi.org/10.1021/acs.macromol.1c01406},
	doi = {10.1021/acs.macromol.1c01406},
	number = {18},
	urldate = {2024-07-02},
	journal = {Macromolecules},
	author = {Saito, Masayuki and Ito, Kohzo and Yokoyama, Hideaki},
	month = sep,
	year = {2021},
	note = {Publisher: American Chemical Society},
	pages = {8538--8547},
}

@article{li_alternating_2019,
	title = {Alternating crystalline lamellar structures from thermodynamically miscible poly(e-caprolactone) {H}/{D} blends},
	volume = {175},
	issn = {0032-3861},
	url = {https://www.sciencedirect.com/science/article/pii/S003238611930480X},
	doi = {10.1016/j.polymer.2019.05.055},
	urldate = {2024-07-02},
	journal = {Polymer},
	author = {Li, Lengwan and Arras, Matthias M. L. and Li, Tianyu and Li, Wei and Chang, Dongsook and Keum, Jong K. and Bonnesen, Peter V. and Qian, Shuo and Peng, Xiangfang and Lee, Byeongdu and Hong, Kunlun},
	month = jun,
	year = {2019},
	pages = {320--328},
}

@article{upadhya_automation_2021,
	title = {Automation and data-driven design of polymer therapeutics},
	volume = {171},
	issn = {0169-409X},
	url = {https://www.sciencedirect.com/science/article/pii/S0169409X20302180},
	doi = {10.1016/j.addr.2020.11.009},
	urldate = {2024-07-02},
	journal = {Advanced Drug Delivery Reviews},
	author = {Upadhya, Rahul and Kosuri, Shashank and Tamasi, Matthew and Meyer, Travis A. and Atta, Supriya and Webb, Michael A. and Gormley, Adam J.},
	month = apr,
	year = {2021},
	pages = {1--28},
}

@article{liang_machine-learning-assisted_2021,
	title = {Machine-learning-assisted low dielectric constant polymer discovery},
	volume = {5},
	issn = {2052-1537},
	url = {https://pubs.rsc.org/en/content/articlelanding/2021/qm/d0qm01093f},
	doi = {10.1039/D0QM01093F},
	language = {en},
	number = {10},
	urldate = {2024-07-02},
	journal = {Materials Chemistry Frontiers},
	author = {Liang, Jiechun and Xu, Shangqian and Hu, Linfeng and Zhao, Yu and Zhu, Xi},
	month = may,
	year = {2021},
	note = {Publisher: The Royal Society of Chemistry},
	pages = {3823--3829},
}

@article{xu_machine_2021,
	title = {Machine {Learning} {Aided} {Design} of {Polymer} with {Targeted} {Band} {Gap} {Based} on {DFT} {Computation}},
	volume = {125},
	issn = {1520-6106},
	url = {https://doi.org/10.1021/acs.jpcb.0c08674},
	doi = {10.1021/acs.jpcb.0c08674},
	number = {2},
	urldate = {2024-07-02},
	journal = {The Journal of Physical Chemistry B},
	author = {Xu, Pengcheng and Lu, Tian and Ju, Lifei and Tian, Lumin and Li, Minjie and Lu, Wencong},
	month = jan,
	year = {2021},
	note = {Publisher: American Chemical Society},
	pages = {601--611},
}

@article{chen_predicting_2021,
	title = {Predicting {Polymers}’ {Glass} {Transition} {Temperature} by a {Chemical} {Language} {Processing} {Model}},
	volume = {13},
	copyright = {http://creativecommons.org/licenses/by/3.0/},
	issn = {2073-4360},
	url = {https://www.mdpi.com/2073-4360/13/11/1898},
	doi = {10.3390/polym13111898},
	language = {en},
	number = {11},
	urldate = {2024-07-02},
	journal = {Polymers},
	author = {Chen, Guang and Tao, Lei and Li, Ying},
	month = jan,
	year = {2021},
	note = {Number: 11
Publisher: Multidisciplinary Digital Publishing Institute},
	pages = {1898},
}

@article{xu_accurate_2008,
	title = {Accurate {Prediction} of $\theta$ ({Lower} {Critical} {Solution} {Temperature}) in {Polymer} {Solutions} {Based} on {3D} {Descriptors} and {Artificial} {Neural} {Networks}},
	volume = {17},
	copyright = {Copyright © 2008 WILEY-VCH Verlag GmbH \& Co. KGaA, Weinheim},
	issn = {1521-3919},
	url = {https://onlinelibrary.wiley.com/doi/abs/10.1002/mats.200700067},
	doi = {10.1002/mats.200700067},
	language = {en},
	number = {2-3},
	urldate = {2024-07-02},
	journal = {Macromolecular Theory and Simulations},
	author = {Xu, Jie and Chen, Biao and Liang, Hao},
	year = {2008},
	note = {\_eprint: https://onlinelibrary.wiley.com/doi/pdf/10.1002/mats.200700067},
	pages = {109--120},
}

@article{macleod_self-driving_2020,
	title = {Self-driving laboratory for accelerated discovery of thin-film materials},
	volume = {6},
	url = {https://www.science.org/doi/10.1126/sciadv.aaz8867},
	doi = {10.1126/sciadv.aaz8867},
	number = {20},
	urldate = {2024-07-02},
	journal = {Science Advances},
	author = {MacLeod, B. P. and Parlane, F. G. L. and Morrissey, T. D. and Häse, F. and Roch, L. M. and Dettelbach, K. E. and Moreira, R. and Yunker, L. P. E. and Rooney, M. B. and Deeth, J. R. and Lai, V. and Ng, G. J. and Situ, H. and Zhang, R. H. and Elliott, M. S. and Haley, T. H. and Dvorak, D. J. and Aspuru-Guzik, A. and Hein, J. E. and Berlinguette, C. P.},
	month = may,
	year = {2020},
	note = {Publisher: American Association for the Advancement of Science},
	pages = {eaaz8867},
}

@article{chen_polymer_2021,
	title = {Polymer informatics: {Current} status and critical next steps},
	volume = {144},
	issn = {0927-796X},
	shorttitle = {Polymer informatics},
	url = {https://www.sciencedirect.com/science/article/pii/S0927796X2030053X},
	doi = {10.1016/j.mser.2020.100595},
	urldate = {2024-07-02},
	journal = {Materials Science and Engineering: R: Reports},
	author = {Chen, Lihua and Pilania, Ghanshyam and Batra, Rohit and Huan, Tran Doan and Kim, Chiho and Kuenneth, Christopher and Ramprasad, Rampi},
	month = apr,
	year = {2021},
	pages = {100595},
}

@article{patra_data-driven_2022,
	title = {Data-{Driven} {Methods} for {Accelerating} {Polymer} {Design}},
	volume = {2},
	url = {https://doi.org/10.1021/acspolymersau.1c00035},
	doi = {10.1021/acspolymersau.1c00035},
	number = {1},
	urldate = {2024-07-02},
	journal = {ACS Polymers Au},
	author = {Patra, Tarak K.},
	month = feb,
	year = {2022},
	note = {Publisher: American Chemical Society},
	pages = {8--26},
}

@article{day_navigating_2023,
	title = {Navigating the {Expansive} {Landscapes} of {Soft} {Materials}: {A} {User} {Guide} for {High}-{Throughput} {Workflows}},
	volume = {3},
	shorttitle = {Navigating the {Expansive} {Landscapes} of {Soft} {Materials}},
	url = {https://doi.org/10.1021/acspolymersau.3c00025},
	doi = {10.1021/acspolymersau.3c00025},
	number = {6},
	urldate = {2024-07-02},
	journal = {ACS Polymers Au},
	author = {Day, Erin C. and Chittari, Supraja S. and Bogen, Matthew P. and Knight, Abigail S.},
	month = dec,
	year = {2023},
	note = {Publisher: American Chemical Society},
	pages = {406--427},
}

@article{schuett_application_2024,
	title = {Application of {Digital} {Methods} in {Polymer} {Science} and {Engineering}},
	volume = {34},
	copyright = {© 2023 The Authors. Advanced Functional Materials published by Wiley-VCH GmbH},
	issn = {1616-3028},
	url = {https://onlinelibrary.wiley.com/doi/abs/10.1002/adfm.202309844},
	doi = {10.1002/adfm.202309844},
	language = {en},
	number = {8},
	urldate = {2024-07-02},
	journal = {Advanced Functional Materials},
	author = {Schuett, Timo and Endres, Patrick and Standau, Tobias and Zechel, Stefan and Albuquerque, Rodrigo Q. and Brütting, Christian and Ruckdäschel, Holger and Schubert, Ulrich S.},
	year = {2024},
	note = {\_eprint: https://onlinelibrary.wiley.com/doi/pdf/10.1002/adfm.202309844},
	pages = {2309844},
}

@article{jordan_machine_2015,
	title = {Machine learning: {Trends}, perspectives, and prospects},
	volume = {349},
	shorttitle = {Machine learning},
	url = {https://www.science.org/doi/10.1126/science.aaa8415},
	doi = {10.1126/science.aaa8415},
	number = {6245},
	urldate = {2024-07-02},
	journal = {Science},
	author = {Jordan, M. I. and Mitchell, T. M.},
	month = jul,
	year = {2015},
	note = {Publisher: American Association for the Advancement of Science},
	pages = {255--260},
}

@article{castillo-boton_machine_2022,
	title = {Machine learning regression and classification methods for fog events prediction},
	volume = {272},
	issn = {0169-8095},
	url = {https://www.sciencedirect.com/science/article/pii/S0169809522001430},
	doi = {10.1016/j.atmosres.2022.106157},
	urldate = {2024-07-02},
	journal = {Atmospheric Research},
	author = {Castillo-Botón, C. and Casillas-Pérez, D. and Casanova-Mateo, C. and Ghimire, S. and Cerro-Prada, E. and Gutierrez, P. A. and Deo, R. C. and Salcedo-Sanz, S.},
	month = jul,
	year = {2022},
	pages = {106157},
}

@article{wei_machine_2019,
	title = {Machine learning in materials science},
	volume = {1},
	copyright = {© 2019 The Authors. InfoMat published by John Wiley \& Sons Australia, Ltd on behalf of UESTC.},
	issn = {2567-3165},
	url = {https://onlinelibrary.wiley.com/doi/abs/10.1002/inf2.12028},
	doi = {10.1002/inf2.12028},
	language = {en},
	number = {3},
	urldate = {2024-07-02},
	journal = {InfoMat},
	author = {Wei, Jing and Chu, Xuan and Sun, Xiang-Yu and Xu, Kun and Deng, Hui-Xiong and Chen, Jigen and Wei, Zhongming and Lei, Ming},
	year = {2019},
	note = {\_eprint: https://onlinelibrary.wiley.com/doi/pdf/10.1002/inf2.12028},
	pages = {338--358},
}

@article{ma_machine-learning-assisted_2023,
	title = {Machine-{Learning}-{Assisted} {Understanding} of {Polymer} {Nanocomposites} {Composition}–{Property} {Relationship}: {A} {Case} {Study} of {NanoMine} {Database}},
	volume = {56},
	issn = {0024-9297},
	shorttitle = {Machine-{Learning}-{Assisted} {Understanding} of {Polymer} {Nanocomposites} {Composition}–{Property} {Relationship}},
	url = {https://doi.org/10.1021/acs.macromol.2c02249},
	doi = {10.1021/acs.macromol.2c02249},
	number = {11},
	urldate = {2024-07-02},
	journal = {Macromolecules},
	author = {Ma, Boran and Finan, Nicholas J. and Jany, David and Deagen, Michael E. and Schadler, Linda S. and Brinson, L. Catherine},
	month = jun,
	year = {2023},
	note = {Publisher: American Chemical Society},
	pages = {3945--3953},
}

@article{lu_accelerated_2018,
	title = {Accelerated discovery of stable lead-free hybrid organic-inorganic perovskites via machine learning},
	volume = {9},
	copyright = {2018 The Author(s)},
	issn = {2041-1723},
	url = {https://www.nature.com/articles/s41467-018-05761-w},
	doi = {10.1038/s41467-018-05761-w},
	language = {en},
	number = {1},
	urldate = {2024-07-02},
	journal = {Nature Communications},
	author = {Lu, Shuaihua and Zhou, Qionghua and Ouyang, Yixin and Guo, Yilv and Li, Qiang and Wang, Jinlan},
	month = aug,
	year = {2018},
	note = {Publisher: Nature Publishing Group},
	pages = {3405},
}

@article{oliynyk_high-throughput_2016,
	title = {High-{Throughput} {Machine}-{Learning}-{Driven} {Synthesis} of {Full}-{Heusler} {Compounds}},
	volume = {28},
	issn = {0897-4756},
	url = {https://doi.org/10.1021/acs.chemmater.6b02724},
	doi = {10.1021/acs.chemmater.6b02724},
	number = {20},
	urldate = {2024-07-02},
	journal = {Chemistry of Materials},
	author = {Oliynyk, Anton O. and Antono, Erin and Sparks, Taylor D. and Ghadbeigi, Leila and Gaultois, Michael W. and Meredig, Bryce and Mar, Arthur},
	month = oct,
	year = {2016},
	note = {Publisher: American Chemical Society},
	pages = {7324--7331},
}

@article{fridgeirsdottir_multiple_2018,
	title = {Multiple {Linear} {Regression} {Modeling} {To} {Predict} the {Stability} of {Polymer}–{Drug} {Solid} {Dispersions}: {Comparison} of the {Effects} of {Polymers} and {Manufacturing} {Methods} on {Solid} {Dispersion} {Stability}},
	volume = {15},
	issn = {1543-8384},
	shorttitle = {Multiple {Linear} {Regression} {Modeling} {To} {Predict} the {Stability} of {Polymer}–{Drug} {Solid} {Dispersions}},
	url = {https://doi.org/10.1021/acs.molpharmaceut.8b00021},
	doi = {10.1021/acs.molpharmaceut.8b00021},
	number = {5},
	urldate = {2024-07-02},
	journal = {Molecular Pharmaceutics},
	author = {Fridgeirsdottir, Gudrun A. and Harris, Robert J. and Dryden, Ian L. and Fischer, Peter M. and Roberts, Clive J.},
	month = may,
	year = {2018},
	note = {Publisher: American Chemical Society},
	pages = {1826--1841},
}

@article{tao_benchmarking_2021,
	title = {Benchmarking {Machine} {Learning} {Models} for {Polymer} {Informatics}: {An} {Example} of {Glass} {Transition} {Temperature}},
	volume = {61},
	issn = {1549-9596},
	shorttitle = {Benchmarking {Machine} {Learning} {Models} for {Polymer} {Informatics}},
	url = {https://doi.org/10.1021/acs.jcim.1c01031},
	doi = {10.1021/acs.jcim.1c01031},
	number = {11},
	urldate = {2024-07-02},
	journal = {Journal of Chemical Information and Modeling},
	author = {Tao, Lei and Varshney, Vikas and Li, Ying},
	month = nov,
	year = {2021},
	note = {Publisher: American Chemical Society},
	pages = {5395--5413},
}

@article{kim_comparison_2021,
	title = {Comparison between {Multiple} {Regression} {Analysis}, {Polynomial} {Regression} {Analysis}, and an {Artificial} {Neural} {Network} for {Tensile} {Strength} {Prediction} of {BFRP} and {GFRP}},
	volume = {14},
	copyright = {http://creativecommons.org/licenses/by/3.0/},
	issn = {1996-1944},
	url = {https://www.mdpi.com/1996-1944/14/17/4861},
	doi = {10.3390/ma14174861},
	language = {en},
	number = {17},
	urldate = {2024-07-02},
	journal = {Materials},
	author = {Kim, Younghwan and Oh, Hongseob},
	month = jan,
	year = {2021},
	note = {Number: 17
Publisher: Multidisciplinary Digital Publishing Institute},
	pages = {4861},
}

@article{banerjee_mathematical_2022,
	title = {Mathematical regression models for rheological behavior of interaction between polymer-surfactant binary mixtures and electrolytes},
	volume = {43},
	issn = {0193-2691},
	url = {https://doi.org/10.1080/01932691.2020.1857261},
	doi = {10.1080/01932691.2020.1857261},
	number = {9},
	urldate = {2024-07-02},
	journal = {Journal of Dispersion Science and Technology},
	author = {Banerjee, Tandrima and Samanta, Abhijit and Mandal, Ajay},
	month = jul,
	year = {2022},
	note = {Publisher: Taylor \& Francis
\_eprint: https://doi.org/10.1080/01932691.2020.1857261},
	pages = {1333--1345},
}

@article{ethier_predicting_2022,
	title = {Predicting {Phase} {Behavior} of {Linear} {Polymers} in {Solution} {Using} {Machine} {Learning}},
	volume = {55},
	issn = {0024-9297},
	url = {https://doi.org/10.1021/acs.macromol.2c00245},
	doi = {10.1021/acs.macromol.2c00245},
	number = {7},
	urldate = {2024-07-02},
	journal = {Macromolecules},
	author = {Ethier, Jeffrey G. and Casukhela, Rohan K. and Latimer, Joshua J. and Jacobsen, Matthew D. and Rasin, Boris and Gupta, Maneesh K. and Baldwin, Luke A. and Vaia, Richard A.},
	month = apr,
	year = {2022},
	note = {Publisher: American Chemical Society},
	pages = {2691--2702},
}

@article{kim_polymer_2021,
	title = {Polymer design using genetic algorithm and machine learning},
	volume = {186},
	issn = {0927-0256},
	url = {https://www.sciencedirect.com/science/article/pii/S0927025620305589},
	doi = {10.1016/j.commatsci.2020.110067},
	urldate = {2024-07-02},
	journal = {Computational Materials Science},
	author = {Kim, Chiho and Batra, Rohit and Chen, Lihua and Tran, Huan and Ramprasad, Rampi},
	month = jan,
	year = {2021},
	pages = {110067},
}

@article{sha_machine_2021,
	title = {Machine learning in polymer informatics},
	volume = {3},
	copyright = {© 2021 The Authors. InfoMat published by UESTC and John Wiley \& Sons Australia, Ltd},
	issn = {2567-3165},
	url = {https://onlinelibrary.wiley.com/doi/abs/10.1002/inf2.12167},
	doi = {10.1002/inf2.12167},
	language = {en},
	number = {4},
	urldate = {2024-07-02},
	journal = {InfoMat},
	author = {Sha, Wuxin and Li, Yan and Tang, Shun and Tian, Jie and Zhao, Yuming and Guo, Yaqing and Zhang, Weixin and Zhang, Xinfang and Lu, Songfeng and Cao, Yuan-Cheng and Cheng, Shijie},
	year = {2021},
	note = {\_eprint: https://onlinelibrary.wiley.com/doi/pdf/10.1002/inf2.12167},
	pages = {353--361},
}

@article{struble_prospective_2024,
	title = {A prospective on machine learning challenges, progress, and potential in polymer science},
	issn = {2159-6867},
	url = {https://doi.org/10.1557/s43579-024-00587-8},
	doi = {10.1557/s43579-024-00587-8},
	language = {en},
	urldate = {2024-07-02},
	journal = {MRS Communications},
	author = {Struble, Daniel C. and Lamb, Bradley G. and Ma, Boran},
	month = jul,
	year = {2024},
}

@article{eyke_toward_2021,
	series = {Special {Issue}: {Machine} {Learning} for {Molecules} and {Materials}},
	title = {Toward {Machine} {Learning}-{Enhanced} {High}-{Throughput} {Experimentation}},
	volume = {3},
	issn = {2589-5974},
	url = {https://www.sciencedirect.com/science/article/pii/S2589597420303117},
	doi = {10.1016/j.trechm.2020.12.001},
	number = {2},
	urldate = {2024-07-02},
	journal = {Trends in Chemistry},
	author = {Eyke, Natalie S. and Koscher, Brent A. and Jensen, Klavs F.},
	month = feb,
	year = {2021},
	pages = {120--132},
}

@article{weller_roll--roll_2019,
	title = {Roll-to-{Roll} {Scalable} {Production} of {Ordered} {Microdomains} through {Nonvolatile} {Additive} {Solvent} {Annealing} of {Block} {Copolymers}},
	volume = {52},
	issn = {0024-9297},
	url = {https://doi.org/10.1021/acs.macromol.9b00772},
	doi = {10.1021/acs.macromol.9b00772},
	number = {13},
	urldate = {2024-07-02},
	journal = {Macromolecules},
	author = {Weller, Daniel W. and Galuska, Luke and Wang, Weiyu and Ehlenburg, Dakota and Hong, Kunlun and Gu, Xiaodan},
	month = jul,
	year = {2019},
	note = {Publisher: American Chemical Society},
	pages = {5026--5032},
}

@article{gu_situ_2014,
	title = {An {In} {Situ} {Grazing} {Incidence} {X}-{Ray} {Scattering} {Study} of {Block} {Copolymer} {Thin} {Films} {During} {Solvent} {Vapor} {Annealing}},
	volume = {26},
	copyright = {© 2013 WILEY-VCH Verlag GmbH \& Co. KGaA, Weinheim},
	issn = {1521-4095},
	url = {https://onlinelibrary.wiley.com/doi/abs/10.1002/adma.201302562},
	doi = {10.1002/adma.201302562},
	number = {2},
	urldate = {2024-07-02},
	journal = {Advanced Materials},
	author = {Gu, Xiaodan and Gunkel, Ilja and Hexemer, Alexander and Gu, Weiyin and Russell, Thomas P.},
	year = {2014},
	note = {\_eprint: https://onlinelibrary.wiley.com/doi/pdf/10.1002/adma.201302562},
	pages = {273--281},
}

@article{carmona_structure_2021,
	title = {Structure evolution during phase separation in spin-coated ethylcellulose/hydroxypropylcellulose films},
	volume = {17},
	issn = {1744-6848},
	url = {https://pubs.rsc.org/en/content/articlelanding/2021/sm/d1sm00044f},
	doi = {10.1039/D1SM00044F},
	language = {en},
	number = {14},
	urldate = {2024-07-02},
	journal = {Soft Matter},
	author = {Carmona, Pierre and Röding, Magnus and Särkkä, Aila and Corswant, Christian von and Olsson, Eva and Lorén, Niklas},
	month = apr,
	year = {2021},
	note = {Publisher: The Royal Society of Chemistry},
	pages = {3913--3922},
}

@inproceedings{zhao_mlexchange_2022,
	title = {{MLExchange}: {A} web-based platform enabling exchangeable machine learning workflows for scientific studies},
	shorttitle = {{MLExchange}},
	url = {http://arxiv.org/abs/2208.09751},
	doi = {10.1109/XLOOP56614.2022.00007},
	urldate = {2024-07-02},
	booktitle = {2022 4th {Annual} {Workshop} on {Extreme}-scale {Experiment}-in-the-{Loop} {Computing} ({XLOOP})},
	author = {Zhao, Zhuowen and Chavez, Tanny and Holman, Elizabeth A. and Hao, Guanhua and Green, Adam and Krishnan, Harinarayan and McReynolds, Dylan and Pandolfi, Ronald and Roberts, Eric J. and Zwart, Petrus H. and Yanxon, Howard and Schwarz, Nicholas and Sankaranarayanan, Subramanian and Kalinin, Sergei V. and Mehta, Apurva and Campbell, Stuart and Hexemer, Alexander},
	month = nov,
	year = {2022},
	note = {arXiv:2208.09751 [cs]},
	pages = {10--15},
}

@article{pedregosa_scikit-learn_2018,
	title = {Scikit-learn: {Machine} {Learning} in {Python}},
	shorttitle = {Scikit-learn},
	url = {http://arxiv.org/abs/1201.0490},
	doi = {10.48550/arXiv.1201.0490},
	urldate = {2024-07-02},
	journal = {arXiv},
	author = {Pedregosa, Fabian and Varoquaux, Gaël and Gramfort, Alexandre and Michel, Vincent and Thirion, Bertrand and Grisel, Olivier and Blondel, Mathieu and Müller, Andreas and Nothman, Joel and Louppe, Gilles and Prettenhofer, Peter and Weiss, Ron and Dubourg, Vincent and Vanderplas, Jake and Passos, Alexandre and Cournapeau, David and Brucher, Matthieu and Perrot, Matthieu and Duchesnay, Édouard},
	month = jun,
	year = {2018},
	note = {accessed 2026-01-05},
}

@article{simonyan_very_2015,
	title = {Very {Deep} {Convolutional} {Networks} for {Large}-{Scale} {Image} {Recognition}},
	url = {http://arxiv.org/abs/1409.1556},
	doi = {10.48550/arXiv.1409.1556},
	urldate = {2024-07-02},
	journal = {arXiv},
	author = {Simonyan, Karen and Zisserman, Andrew},
	month = apr,
	year = {2015},
	note = {accessed 2026-01-05},
}

@article{lundberg_unified_2017,
	title = {A {Unified} {Approach} to {Interpreting} {Model} {Predictions}},
	url = {http://arxiv.org/abs/1705.07874},
	doi = {10.48550/arXiv.1705.07874},
	urldate = {2024-07-02},
	journal = {arXiv},
	author = {Lundberg, Scott and Lee, Su-In},
	month = nov,
	year = {2017},
	note = {accessed 2026-01-05},
}

@article{lundberg_local_2020,
	title = {From local explanations to global understanding with explainable {AI} for trees},
	volume = {2},
	copyright = {2020 The Author(s), under exclusive licence to Springer Nature Limited},
	issn = {2522-5839},
	url = {https://www.nature.com/articles/s42256-019-0138-9},
	doi = {10.1038/s42256-019-0138-9},
	language = {en},
	number = {1},
	urldate = {2024-07-02},
	journal = {Nature Machine Intelligence},
	author = {Lundberg, Scott M. and Erion, Gabriel and Chen, Hugh and DeGrave, Alex and Prutkin, Jordan M. and Nair, Bala and Katz, Ronit and Himmelfarb, Jonathan and Bansal, Nisha and Lee, Su-In},
	month = jan,
	year = {2020},
	note = {Publisher: Nature Publishing Group},
	pages = {56--67},
}

@article{chang_accelerating_2019,
	title = {Accelerating {Neutron} {Scattering} {Data} {Collection} and {Experiments} {Using} {AI} {Deep} {Super}-{Resolution} {Learning}},
	url = {http://arxiv.org/abs/1904.08450},
	doi = {10.48550/arXiv.1904.08450},
	urldate = {2024-07-02},
	journal = {arXiv},
	author = {Chang, Ming-Ching and Wei, Yi and Chen, Wei-Ren and Do, Changwoo},
	month = may,
	year = {2019},
    note = {accessed 2026-01-05},
}

@article{lorenzoni_assessing_2015,
	title = {Assessing the {Local} {Nanomechanical} {Properties} of {Self}-{Assembled} {Block} {Copolymer} {Thin} {Films} by {Peak} {Force} {Tapping}},
	volume = {31},
	issn = {0743-7463},
	url = {https://doi.org/10.1021/acs.langmuir.5b02595},
	doi = {10.1021/acs.langmuir.5b02595},
	number = {42},
	urldate = {2024-07-02},
	journal = {Langmuir},
	author = {Lorenzoni, Matteo and Evangelio, Laura and Verhaeghe, Sophie and Nicolet, Célia and Navarro, Christophe and Pérez-Murano, Francesc},
	month = oct,
	year = {2015},
	note = {Publisher: American Chemical Society},
	pages = {11630--11638},
}

@article{mansky_controlling_1997,
	title = {Controlling {Polymer}-{Surface} {Interactions} with {Random} {Copolymer} {Brushes}},
	volume = {275},
	url = {https://www.science.org/doi/10.1126/science.275.5305.1458},
	doi = {10.1126/science.275.5305.1458},
	number = {5305},
	urldate = {2024-07-02},
	journal = {Science},
	author = {Mansky, P. and Liu, Y. and Huang, E. and Russell, T. P. and Hawker, C.},
	month = mar,
	year = {1997},
	note = {Publisher: American Association for the Advancement of Science},
	pages = {1458--1460},
}

@article{brassat_nanoscale_2020,
	title = {Nanoscale {Block} {Copolymer} {Self}-{Assembly} and {Microscale} {Polymer} {Film} {Dewetting}: {Progress} in {Understanding} the {Role} of {Interfacial} {Energies} in the {Formation} of {Hierarchical} {Nanostructures}},
	volume = {7},
	copyright = {© 2019 The Authors. Published by WILEY-VCH Verlag GmbH \& Co. KGaA, Weinheim},
	issn = {2196-7350},
	shorttitle = {Nanoscale {Block} {Copolymer} {Self}-{Assembly} and {Microscale} {Polymer} {Film} {Dewetting}},
	url = {https://onlinelibrary.wiley.com/doi/abs/10.1002/admi.201901565},
	doi = {10.1002/admi.201901565},
	language = {en},
	number = {5},
	urldate = {2024-07-02},
	journal = {Advanced Materials Interfaces},
	author = {Brassat, Katharina and Lindner, Jörg K. N.},
	year = {2020},
	note = {\_eprint: https://onlinelibrary.wiley.com/doi/pdf/10.1002/admi.201901565},
	pages = {1901565},
}

@article{hou_applications_2019,
	title = {The {Applications} of {Polymers} in {Solar} {Cells}: {A} {Review}},
	volume = {11},
	copyright = {http://creativecommons.org/licenses/by/3.0/},
	issn = {2073-4360},
	shorttitle = {The {Applications} of {Polymers} in {Solar} {Cells}},
	url = {https://www.mdpi.com/2073-4360/11/1/143},
	doi = {10.3390/polym11010143},
	language = {en},
	number = {1},
	urldate = {2024-07-02},
	journal = {Polymers},
	author = {Hou, Wenjing and Xiao, Yaoming and Han, Gaoyi and Lin, Jeng-Yu},
	month = jan,
	year = {2019},
	note = {Number: 1
Publisher: Multidisciplinary Digital Publishing Institute},
	pages = {143},
}

@article{leibler_theory_1980,
	title = {Theory of {Microphase} {Separation} in {Block} {Copolymers}},
	volume = {13},
	issn = {0024-9297},
	url = {https://doi.org/10.1021/ma60078a047},
	doi = {10.1021/ma60078a047},
	number = {6},
	urldate = {2024-07-02},
	journal = {Macromolecules},
	author = {Leibler, Ludwik},
	month = nov,
	year = {1980},
	note = {Publisher: American Chemical Society},
	pages = {1602--1617},
}

@article{matsen_unifying_1996,
	title = {Unifying {Weak}- and {Strong}-{Segregation} {Block} {Copolymer} {Theories}},
	volume = {29},
	issn = {0024-9297},
	url = {https://doi.org/10.1021/ma951138i},
	doi = {10.1021/ma951138i},
	number = {4},
	urldate = {2024-07-02},
	journal = {Macromolecules},
	author = {Matsen, M. W. and Bates, F. S.},
	month = jan,
	year = {1996},
	note = {Publisher: American Chemical Society},
	pages = {1091--1098},
}

@article{liu_unraveling_2018,
	title = {Unraveling the {Main} {Chain} and {Side} {Chain} {Effects} on {Thin} {Film} {Morphology} and {Charge} {Transport} in {Quinoidal} {Conjugated} {Polymers}},
	volume = {28},
	copyright = {© 2018 WILEY-VCH Verlag GmbH \& Co. KGaA, Weinheim},
	issn = {1616-3028},
	url = {https://onlinelibrary.wiley.com/doi/abs/10.1002/adfm.201801874},
	doi = {10.1002/adfm.201801874},
	language = {en},
	number = {31},
	urldate = {2024-07-02},
	journal = {Advanced Functional Materials},
	author = {Liu, Xuncheng and He, Bo and Garzón-Ruiz, Andrés and Navarro, Amparo and Chen, Teresa L. and Kolaczkowski, Matthew A. and Feng, Shizhen and Zhang, Lianjie and Anderson, Christopher A. and Chen, Junwu and Liu, Yi},
	year = {2018},
	note = {\_eprint: https://onlinelibrary.wiley.com/doi/pdf/10.1002/adfm.201801874},
	pages = {1801874},
}

@article{murphy_automated_2015,
	title = {Automated {Defect} and {Correlation} {Length} {Analysis} of {Block} {Copolymer} {Thin} {Film} {Nanopatterns}},
	volume = {10},
	issn = {1932-6203},
	url = {https://journals.plos.org/plosone/article?id=10.1371/journal.pone.0133088},
	doi = {10.1371/journal.pone.0133088},
	language = {en},
	number = {7},
	urldate = {2024-07-12},
	journal = {PLOS ONE},
	author = {Murphy, Jeffrey N. and Harris, Kenneth D. and Buriak, Jillian M.},
	month = jul,
	year = {2015},
	note = {Publisher: Public Library of Science},
	pages = {e0133088},
}

@article{ogieglo_situ_2015,
	series = {Topical {Issue} on {Polymer} {Physics}},
	title = {In situ ellipsometry studies on swelling of thin polymer films: {A} review},
	volume = {42},
	issn = {0079-6700},
	shorttitle = {In situ ellipsometry studies on swelling of thin polymer films},
	url = {https://www.sciencedirect.com/science/article/pii/S0079670014001063},
	doi = {10.1016/j.progpolymsci.2014.09.004},
	urldate = {2024-07-12},
	journal = {Progress in Polymer Science},
	author = {Ogieglo, Wojciech and Wormeester, Herbert and Eichhorn, Klaus-Jochen and Wessling, Matthias and Benes, Nieck E.},
	month = mar,
	year = {2015},
	pages = {42--78},
}

@misc{wada_wkentarolabelme_2021,
	title = {LabelMe: Image Polygonal Annotation with Python},
	shorttitle = {wkentaro/labelme},
	url = {https://zenodo.org/records/5711226},
	urldate = {2025-05-01},
	publisher = {Zenodo},
	author = {Wada, Kentaro and mpitid and Buijs, Martijn and N, Zhang Ch and なるみ and Kubovčík, Bc Martin and Myczko, Alex and latentix and Zhu, Lingjie and Yamaguchi, Naoya and Fujii, Shohei and iamgd67 and IlyaOvodov and Patel, Akshar and Clauss, Christian and Kuroiwa, Eisoku and Iyengar, Roger and Shilin, Sergei and Malygina, Tanya and Kawaharazuka, Kento and Engelberts, Jonne and J, Aleksi and AlexMa and Song, Changwoo and Charlie and Rose, Daniel and Livingstone, Douglas and Doug and Erik and Toft, Henrik},
	month = nov,
	year = {2021},
	doi = {10.5281/zenodo.5711226},
    note = {accessed 2026-01-05},
}

@article{bourne_hexagonal_2014,
	title = {Hexagonal {Patterns} in a {Simplified} {Model} for {Block} {Copolymers}},
	volume = {74},
	issn = {0036-1399},
	url = {https://epubs.siam.org/doi/abs/10.1137/130922732},
	doi = {10.1137/130922732},
	number = {5},
	urldate = {2025-05-01},
	journal = {SIAM Journal on Applied Mathematics},
	author = {Bourne, D. P. and Peletier, M. A. and Roper, S. M.},
	month = jan,
	year = {2014},
	note = {Publisher: Society for Industrial and Applied Mathematics},
	pages = {1315--1337},
}

@article{hammond_adjustment_2003,
	title = {Adjustment of {Block} {Copolymer} {Nanodomain} {Sizes} at {Lattice} {Defect} {Sites}},
	volume = {36},
	issn = {0024-9297},
	url = {https://doi.org/10.1021/ma026001o},
	doi = {10.1021/ma026001o},
	number = {23},
	urldate = {2025-05-01},
	journal = {Macromolecules},
	author = {Hammond, Matthew R. and Sides, Scott W. and Fredrickson, Glenn H. and Kramer, Edward J. and Ruokolainen, Janne and Hahn, Stephen F.},
	month = nov,
	year = {2003},
	note = {Publisher: American Chemical Society},
	pages = {8712--8716},
}

@article{beltran-villegas_computational_2019,
	title = {Computational {Reverse}-{Engineering} {Analysis} for {Scattering} {Experiments} on {Amphiphilic} {Block} {Polymer} {Solutions}},
	volume = {141},
	copyright = {https://doi.org/10.15223/policy-029},
	issn = {0002-7863, 1520-5126},
	url = {https://pubs.acs.org/doi/10.1021/jacs.9b08028},
	doi = {10.1021/jacs.9b08028},
	language = {en},
	number = {37},
	urldate = {2025-05-06},
	journal = {Journal of the American Chemical Society},
	author = {Beltran-Villegas, Daniel J. and Wessels, Michiel G. and Lee, Jee Young and Song, Yue and Wooley, Karen L. and Pochan, Darrin J. and Jayaraman, Arthi},
	month = sep,
	year = {2019},
	pages = {14916--14930},
}

@article{wu_machine_2022,
	title = {Machine {Learning}-{Enhanced} {Computational} {Reverse}-{Engineering} {Analysis} for {Scattering} {Experiments} ({CREASE}) for {Analyzing} {Fibrillar} {Structures} in {Polymer} {Solutions}},
	volume = {55},
	copyright = {https://doi.org/10.15223/policy-029},
	issn = {0024-9297, 1520-5835},
	url = {https://pubs.acs.org/doi/10.1021/acs.macromol.2c02165},
	doi = {10.1021/acs.macromol.2c02165},
	language = {en},
	number = {24},
	urldate = {2025-05-06},
	journal = {Macromolecules},
	author = {Wu, Zijie and Jayaraman, Arthi},
	month = dec,
	year = {2022},
	pages = {11076--11091},
}

@article{wessels_machine_2021,
	title = {Machine {Learning} {Enhanced} {Computational} {Reverse} {Engineering} {Analysis} for {Scattering} {Experiments} ({CREASE}) to {Determine} {Structures} in {Amphiphilic} {Polymer} {Solutions}},
	volume = {1},
	url = {https://doi.org/10.1021/acspolymersau.1c00015},
	doi = {10.1021/acspolymersau.1c00015},
	number = {3},
	urldate = {2025-05-06},
	journal = {ACS Polymers Au},
	author = {Wessels, Michiel G. and Jayaraman, Arthi},
	month = dec,
	year = {2021},
	note = {Publisher: American Chemical Society},
	pages = {153--164},
}

@article{shui_rapid_2025,
	title = {Rapid {Assembly} of {Block} {Copolymer} {Thin} {Films} via {Accelerating} the {Swelling} {Process} {During} {Solvent} {Annealing}},
	volume = {17},
	copyright = {http://creativecommons.org/licenses/by/3.0/},
	issn = {2073-4360},
	url = {https://www.mdpi.com/2073-4360/17/9/1242},
	doi = {10.3390/polym17091242},
	language = {en},
	number = {9},
	urldate = {2025-05-06},
	journal = {Polymers},
	author = {Shui, Tian-en and Chang, Tongxin and Wang, Zhe and Huang, Haiying},
	month = jan,
	year = {2025},
	note = {Number: 9
Publisher: Multidisciplinary Digital Publishing Institute},
	pages = {1242},
}

@article{hao_self-assembly_2017,
	title = {Self-{Assembly} in {Block} {Copolymer} {Thin} {Films} upon {Solvent} {Evaporation}: {A} {Simulation} {Study}},
	volume = {50},
	issn = {0024-9297},
	shorttitle = {Self-{Assembly} in {Block} {Copolymer} {Thin} {Films} upon {Solvent} {Evaporation}},
	url = {https://doi.org/10.1021/acs.macromol.7b00200},
	doi = {10.1021/acs.macromol.7b00200},
	number = {11},
	urldate = {2025-05-07},
	journal = {Macromolecules},
	author = {Hao, Jinlong and Wang, Zhan and Wang, Zheng and Yin, Yuhua and Jiang, Run and Li, Baohui and Wang, Qiang},
	month = jun,
	year = {2017},
	note = {Publisher: American Chemical Society},
	pages = {4384--4396},
}

@article{tung_small_2022,
	title = {Small angle scattering of diblock copolymers profiled by machine learning},
	volume = {156},
	issn = {0021-9606, 1089-7690},
	url = {https://pubs.aip.org/jcp/article/156/13/131101/2840953/Small-angle-scattering-of-diblock-copolymers},
	doi = {10.1063/5.0086311},
	language = {en},
	number = {13},
	urldate = {2025-05-08},
	journal = {The Journal of Chemical Physics},
	author = {Tung, Chi-Huan and Chang, Shou-Yi and Chen, Hsin-Lung and Wang, Yangyang and Hong, Kunlun and Carrillo, Jan Michael and Sumpter, Bobby G. and Shinohara, Yuya and Do, Changwoo and Chen, Wei-Ren},
	month = apr,
	year = {2022},
	pages = {131101},
}

@article{ding_deciphering_2025,
	title = {Deciphering the {Scattering} of {Mechanically} {Driven} {Polymers} {Using} {Deep} {Learning}},
	volume = {21},
	issn = {1549-9618},
	url = {https://doi.org/10.1021/acs.jctc.5c00409},
	doi = {10.1021/acs.jctc.5c00409},
	number = {8},
	urldate = {2025-05-08},
	journal = {Journal of Chemical Theory and Computation},
	author = {Ding, Lijie and Tung, Chi-Huan and Sumpter, Bobby G. and Chen, Wei-Ren and Do, Changwoo},
	month = apr,
	year = {2025},
	note = {Publisher: American Chemical Society},
	pages = {4176--4182},
}

@article{kobayashi_machine_2023,
	title = {Machine learning of atomic force microscopy images of organic solar cells},
	volume = {13},
	url = {https://pubs.rsc.org/en/content/articlelanding/2023/ra/d3ra02492j},
	doi = {10.1039/D3RA02492J},
	language = {en},
	number = {22},
	urldate = {2025-05-08},
	journal = {RSC Advances},
	author = {Kobayashi, Yasuhito and Miyake, Yuta and Ishiwari, Fumitaka and Ishiwata, Shintaro and Saeki, Akinori},
	year = {2023},
	note = {Publisher: Royal Society of Chemistry},
	pages = {15107--15113},
}

@article{paruchuri_machine_2024,
	title = {Machine learning for analyzing atomic force microscopy ({AFM}) images generated from polymer blends},
	volume = {3},
	url = {https://pubs.rsc.org/en/content/articlelanding/2024/dd/d4dd00215f},
	doi = {10.1039/D4DD00215F},
	language = {en},
	number = {12},
	urldate = {2025-05-08},
	journal = {Digital Discovery},
	author = {Paruchuri, Aanish and Wang, Yunfei and Gu, Xiaodan and Jayaraman, Arthi},
	year = {2024},
	note = {Publisher: Royal Society of Chemistry},
	pages = {2533--2550},
}

@article{beaucage_automation_2024,
	title = {Automation and {Machine} {Learning} for {Accelerated} {Polymer} {Characterization} and {Development}: {Past}, {Potential}, and a {Path} {Forward}},
	volume = {57},
	issn = {0024-9297},
	shorttitle = {Automation and {Machine} {Learning} for {Accelerated} {Polymer} {Characterization} and {Development}},
	url = {https://doi.org/10.1021/acs.macromol.4c01410},
	doi = {10.1021/acs.macromol.4c01410},
	number = {18},
	urldate = {2025-08-28},
	journal = {Macromolecules},
	author = {Beaucage, Peter A. and Sutherland, Duncan R. and Martin, Tyler B.},
	month = sep,
	year = {2024},
	note = {Publisher: American Chemical Society},
	pages = {8661--8670},
}

@article{doerk_autonomous_2023,
	title = {Autonomous discovery of emergent morphologies in directed self-assembly of block copolymer blends},
	volume = {9},
	url = {https://www.science.org/doi/10.1126/sciadv.add3687},
	doi = {10.1126/sciadv.add3687},
	number = {2},
	urldate = {2025-08-28},
	journal = {Science Advances},
	author = {Doerk, Gregory S. and Stein, Aaron and Bae, Suwon and Noack, Marcus M. and Fukuto, Masafumi and Yager, Kevin G.},
	month = jan,
	year = {2023},
	note = {Publisher: American Association for the Advancement of Science},
	pages = {eadd3687},
}

@article{bendejacq_well-ordered_2002,
	title = {Well-{Ordered} {Microdomain} {Structures} in {Polydisperse} {Poly}(styrene) {Poly}(acrylic acid) {Diblock} {Copolymers} from {Controlled} {Radical} {Polymerization}},
	volume = {35},
	issn = {0024-9297},
	url = {https://doi.org/10.1021/ma020158z},
	doi = {10.1021/ma020158z},
	number = {17},
	urldate = {2025-09-04},
	journal = {Macromolecules},
	author = {Bendejacq, D. and Ponsinet, V. and Joanicot, M. and Loo, Y.-L. and Register, R. A.},
	month = aug,
	year = {2002},
	note = {Publisher: American Chemical Society},
	pages = {6645--6649},
}

@article{chu_morphologies_1995,
	title = {Morphologies of strongly segregated polystyrene-poly(dimethylsiloxane) diblock copolymers},
	volume = {36},
	issn = {0032-3861},
	url = {https://www.sciencedirect.com/science/article/pii/003238619599001B},
	doi = {10.1016/0032-3861(95)99001-B},
	number = {8},
	urldate = {2025-09-04},
	journal = {Polymer},
	author = {Chu, Jennifer H. and Rangarajan, Pratima and Adams, J. LaMonte and Register, Richard A.},
	month = jan,
	year = {1995},
	pages = {1569--1575},
}

@article{smarsly_quantitative_2005,
	title = {Quantitative {SAXS} {Analysis} of {Oriented} {2D} {Hexagonal} {Cylindrical} {Silica} {Mesostructures} in {Thin} {Films} {Obtained} from {Nonionic} {Surfactants}},
	volume = {21},
	issn = {0743-7463},
	url = {https://doi.org/10.1021/la046916r},
	doi = {10.1021/la046916r},
	number = {9},
	urldate = {2025-09-04},
	journal = {Langmuir},
	author = {Smarsly, Bernd and Gibaud, Alain and Ruland, Wilhelm and Sturmayr, Dietmar and Brinker, C. Jeffrey},
	month = apr,
	year = {2005},
	note = {Publisher: American Chemical Society},
	pages = {3858--3866},
}

@article{williamson_design_2022,
	title = {Design of {Experiments} for {Nanocrystal} {Syntheses}: {A} {How}-{To} {Guide} for {Proper} {Implementation}},
	volume = {34},
	issn = {0897-4756},
	shorttitle = {Design of {Experiments} for {Nanocrystal} {Syntheses}},
	url = {https://doi.org/10.1021/acs.chemmater.2c02924},
	doi = {10.1021/acs.chemmater.2c02924},
	number = {22},
	urldate = {2025-09-05},
	journal = {Chemistry of Materials},
	author = {Williamson, Emily M. and Sun, Zhaohong and Mora-Tamez, Lucía and Brutchey, Richard L.},
	month = nov,
	year = {2022},
	note = {Publisher: American Chemical Society},
	pages = {9823--9835},
}

@article{bukys_high-dimensional_2020,
	title = {High-{Dimensional} {Design}-{Of}-{Experiments} {Extracts} {Small}-{Molecule}-{Only} {Induction} {Conditions} for {Dorsal} {Pancreatic} {Endoderm} from {Pluripotency}},
	volume = {23},
	issn = {2589-0042},
	url = {https://www.ncbi.nlm.nih.gov/pmc/articles/PMC7398937/},
	doi = {10.1016/j.isci.2020.101346},
	number = {8},
	urldate = {2025-09-05},
	journal = {iScience},
	author = {Bukys, Michael A. and Mihas, Alexander and Finney, Krystal and Sears, Katie and Trivedi, Divya and Wang, Yong and Oberholzer, Jose and Jensen, Jan},
	month = jul,
	year = {2020},
	pmid = {32745983},
	pmcid = {PMC7398937},
	pages = {101346},
}

@article{cao_how_2018,
	title = {How {To} {Optimize} {Materials} and {Devices} via {Design} of {Experiments} and {Machine} {Learning}: {Demonstration} {Using} {Organic} {Photovoltaics}},
	volume = {12},
	issn = {1936-0851},
	shorttitle = {How {To} {Optimize} {Materials} and {Devices} via {Design} of {Experiments} and {Machine} {Learning}},
	url = {https://doi.org/10.1021/acsnano.8b04726},
	doi = {10.1021/acsnano.8b04726},
	number = {8},
	urldate = {2025-09-05},
	journal = {ACS Nano},
	author = {Cao, Bing and Adutwum, Lawrence A. and Oliynyk, Anton O. and Luber, Erik J. and Olsen, Brian C. and Mar, Arthur and Buriak, Jillian M.},
	month = aug,
	year = {2018},
	note = {Publisher: American Chemical Society},
	pages = {7434--7444},
}

@article{hamley_hexagonal_1993,
	title = {Hexagonal mesophases between lamellae and cylinders in a diblock copolymer melt},
	volume = {26},
	issn = {0024-9297},
	url = {https://doi.org/10.1021/ma00074a018},
	doi = {10.1021/ma00074a018},
	number = {22},
	urldate = {2025-09-09},
	journal = {Macromolecules},
	author = {Hamley, Ian W. and Koppi, Kurt A. and Rosedale, Jeffrey H. and Bates, Frank S. and Almdal, Kristoffer and Mortensen, Kell},
	month = oct,
	year = {1993},
	note = {Publisher: American Chemical Society},
	pages = {5959--5970},
}

@article{wall_practical_2025,
	title = {A {Practical} {Start}-{Up} {Guide} for {Synthetic} {Chemists} to {Implement} {Design} of {Experiments} ({DoE})},
	volume = {15},
	url = {https://doi.org/10.1021/acscatal.5c01626},
	doi = {10.1021/acscatal.5c01626},
	number = {11},
	urldate = {2025-11-19},
	journal = {ACS Catalysis},
	author = {Wall, Brendan J. and Koeritz, Mason T. and Stanley, Levi M. and VanVeller, Brett},
	month = jun,
	year = {2025},
	note = {Publisher: American Chemical Society},
	pages = {8885--8893},
}

@article{albert_systematic_2012,
	title = {Systematic {Study} on the {Effect} of {Solvent} {Removal} {Rate} on the {Morphology} of {Solvent} {Vapor} {Annealed} {ABA} {Triblock} {Copolymer} {Thin} {Films}},
	volume = {6},
	issn = {1936-0851},
	url = {https://doi.org/10.1021/nn203776c},
	doi = {10.1021/nn203776c},
	number = {1},
	urldate = {2025-11-21},
	journal = {ACS Nano},
	author = {Albert, Julie N. L. and Young, Wen-Shiue and Lewis, Ronald L. III and Bogart, Timothy D. and Smith, Jasmine R. and Epps, Thomas H. III},
	month = jan,
	year = {2012},
	note = {Publisher: American Chemical Society},
	pages = {459--466},
}

\end{document}